\documentclass{article}

\PassOptionsToPackage{numbers, compress}{natbib}

 \usepackage[main, final]{neurips_2026}

\usepackage[utf8]{inputenc}
\usepackage[T1]{fontenc}
\usepackage{hyperref}
\usepackage{url}
\usepackage{microtype}
\usepackage{nicefrac}

\usepackage{amsmath}
\usepackage{amssymb}
\usepackage{amsfonts}
\usepackage{mathtools}
\usepackage{amsthm}
\usepackage{bm}

\usepackage{algorithm}
\usepackage{algorithmic}

\usepackage{graphicx}
\usepackage{booktabs}
\usepackage{multirow}
\usepackage{subcaption}
\usepackage{threeparttable}
\usepackage{colortbl}
\usepackage{arydshln}
\usepackage{diagbox}
\usepackage{wrapfig}

\usepackage{xcolor}
\usepackage{color}
\usepackage{soul}
\usepackage{textcomp}

\usepackage[most]{tcolorbox}
\usepackage{varwidth}
\usepackage{tikz}
\usepackage{tikz-qtree}
\usepackage{pgfplots}
\usetikzlibrary{patterns,plotmarks}

\usepackage{enumitem}
\usepackage{listings}
\usepackage{bbding}
\usepackage{pifont}
\usepackage{balance}
\usepackage{flushend}

\newcommand{\D}{\mathcal{D}}
\newcommand{\T}{\mathcal{T}}

\newcommand{\policy}{\pi_{\theta}}
\newcommand{\grad}{\mathbf{g}}

\definecolor{barOrange}{HTML}{F4B183}
\definecolor{barBlue}{HTML}{9DC3E6}
\definecolor{barYellow}{HTML}{FFD966}
\definecolor{barGreen}{HTML}{A9D18E}
\definecolor{barGray}{HTML}{7F7F7F}
\definecolor{labelRed}{HTML}{C00000}
\definecolor{deep_red}{HTML}{C00000}
\definecolor{deep_green}{HTML}{006400}

\newtcolorbox{promptbox}[1]{
  colback=blue!4!white, colframe=blue!40!black,
  title={#1}, fonttitle=\bfseries\small, left=4pt, right=4pt, top=2pt, bottom=2pt}

\newtcolorbox{badbox}{
  colback=red!5!white, colframe=red!50!black, left=4pt, right=4pt, top=2pt, bottom=2pt}
\newtcolorbox{goodbox}{
  colback=green!5!white, colframe=green!50!black, left=4pt, right=4pt, top=2pt, bottom=2pt}

\newcommand{\redvar}[1]{\textcolor{red}{\texttt{[#1]}}}
\newcommand{\blueback}[1]{\colorbox{cyan!18}{\texttt{#1}}}

\newcommand{\mytab}{\hspace{1.5em}}

\newcommand{\tool}{ASTOR}
\newcommand{\moduleone}{Hierarchical Utility-Routed Data Scheduling}
\newcommand{\moduletwo}{Adaptive Utility-Calibrated Policy Optimization}
\newcommand{\blockone}{Dual-Level Utility Estimation}
\newcommand{\blocktwo}{Hierarchical Data Scheduling}
\newcommand{\blockthree}{Multi-Task Coding Reward System}
\newcommand{\blockfour}{Utility-Calibrated Policy Optimization}

\title{\textsc{Astor}: Multi-Task Code Reinforcement Learning via Utility-Driven Coordination}

\author{
\textbf{Yujia Chen}$^{1}$
~~
\textbf{Yang Ye}$^{2}$
~~
\textbf{Xiao Chu}$^{2}$
~~
\textbf{Yuchi Ma}$^{2}$
~~
\textbf{Cuiyun Gao}$^{1}\thanks{Corresponding Author.}$ \\
\texttt{yujiachen@stu.hit.edu.cn}
~~
\texttt{gaocuiyun@hit.edu.cn}
\\
\texttt{\{yeyang14, chuxiao1, mayuchi1\}@huawei.com} \\
\\
$^{1}$ Harbin Institute of Technology, Shenzhen
~~
$^{2}$ Huawei Cloud Computing Technologies Co., Ltd. \\
}

\begin{document}

\maketitle

\begin{abstract}
    Reinforcement learning (RL) with verifiable rewards has proven effective at post-training LLMs for coding, yet deploying separate task-specific specialists incurs costs that scale with the number of tasks, motivating a unified multi-task RL (MTRL) approach.
However, existing MTRL methods treat all coding tasks uniformly, relying on fixed data curricula under a shared optimization strategy, ultimately limiting the effectiveness of multi-task training.
To address these limitations, we propose \textbf{{\tool}}, a multi-t\textbf{AS}k code reinforcement learning framework via u\textbf{T}ility-driven co\textbf{OR}dination. Centered on \textit{task utility}, a signal capturing each task's learning potential and cross-task synergy, {\tool} comprises two coupled modules: \textit{1) \moduleone} module hierarchically allocates training budget and prioritizes informative prompts, steering training toward the most valuable data; and \textit{2) \moduletwo} module dynamically scales per-task KL regularization, matching update constraints to each task's current training state. 
Experiments on two widely-used LLMs across four representative coding tasks demonstrate that {\tool} consistently improves a single model across all tasks, outperforming the best task-specific specialist in average performance across all tasks by 9.0\%--9.5\% and surpassing the strongest MTRL baseline by 7.5\%--12.8\%.


\end{abstract}

\section{Introduction}
\begin{figure}[t]
    \centering
    \begin{subfigure}[t]{0.95\linewidth}
        \centering
        \includegraphics[width=\linewidth]{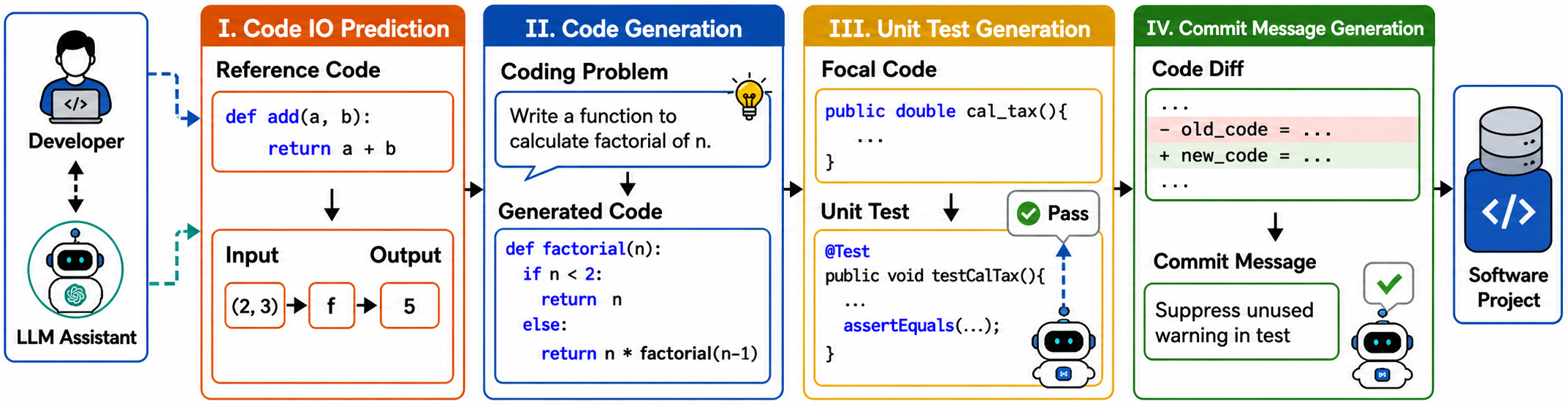}
        \caption{Four representative tasks spanning the software development workflow.}
        \label{fig:development}
    \end{subfigure}
    \hfill
    \begin{subfigure}[t]{\linewidth}
        \centering
        \includegraphics[width=\linewidth]{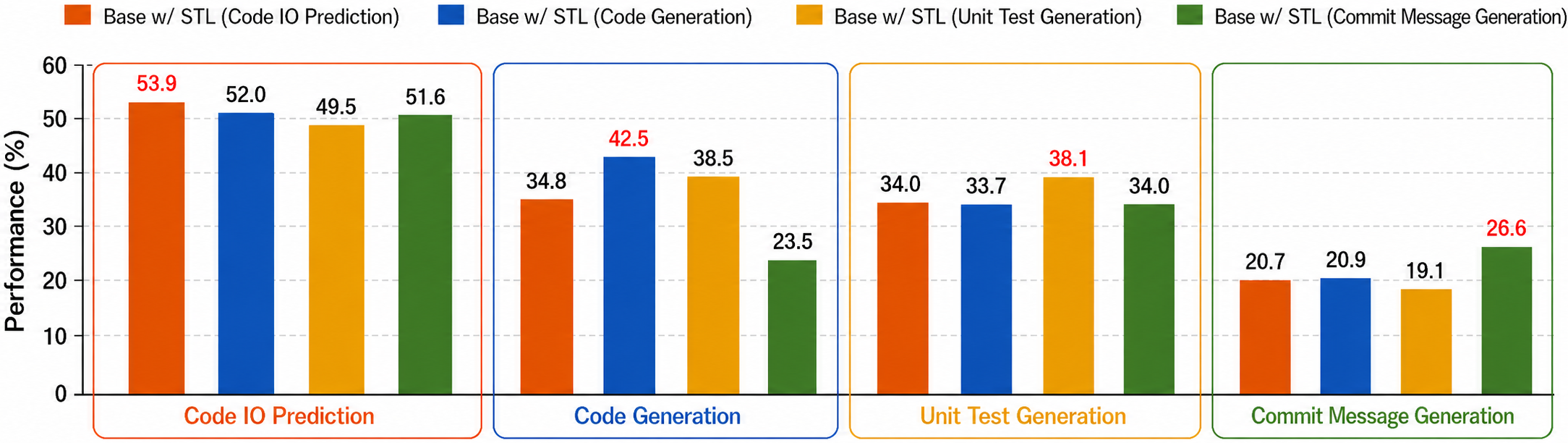}
        \caption{Performance of task-specific RL specialists across the four tasks.}
        \label{fig:performance}
    \end{subfigure}
\caption{Performance of task-specific RL models across four coding tasks. Single-task learning (STL) models perform well on their own tasks but fall short of the corresponding specialists on other tasks.}
    \label{fig:pri_study}
    \vspace{-1em}
\end{figure}

    In the era of large language models (LLMs), AI-assisted coding has become a mainstream development paradigm~\cite{metagpt,se,se2,DBLP:journals/pacmse/ChenYLMG25}. To complete a software project, developers typically need to understand program logic, generate code implementations, validate correctness through unit testing, and record changes via commit messages, as illustrated in Figure~\ref{fig:development}. To enhance the capabilities of LLMs on these tasks, reinforcement learning (RL) has emerged as a powerful post-training technique, since coding tasks naturally provide execution-based reward signals~\cite{intro-RL-code-1, intro-RL-code-2, intro-RL-code-3}.
Following this direction, recent works train task-specific RL specialists tailored to individual stages of the development workflow~\cite{intro-RL-code-4, intro-RL-code-5}. However, deploying separate specialists per task incurs memory and compute costs that scale linearly with task count. Relying on a single specialist offers no remedy either, as no specialist matches the other specialists on their respective tasks (Figure~\ref{fig:performance}).

A natural solution is multi-task reinforcement learning (MTRL)~\cite{intro-MTRL-def,intro-MTRL-Tra-1,intro-MTRL-Tra-2,intro-MTRL-Tra-3,intro-MTRL-Tra-4}, which trains a single LLM jointly across all coding tasks to achieve diverse capabilities without per-task deployment costs.
In LLM post-training, we identify two coupled decisions at each MTRL step: \textit{what to learn}, i.e., scheduling which tasks and prompts to train on, and \textit{how to learn}, i.e., updating the shared policy across heterogeneous tasks. However, existing methods apply uniform strategies to both decisions, ignoring the inherent heterogeneity among coding tasks.


\begin{itemize}[leftmargin=*]

 \item[\(\blacktriangleright\)] For \textbf{what to learn}, existing methods either sample from a mixed task pool or follow a fixed curriculum, relying mainly on task-intrinsic signals~\cite{intro-MTRL-2,intro-MTRL-3}. They ignore inter-task effects; for example, training on code generation may affect unit test generation, as both require program reasoning but emphasize different output behaviors. This leads to suboptimal task schedules.

 \item[\(\blacktriangleright\)] For \textbf{how to learn}, existing methods usually apply the same update strategy to all tasks~\cite{intro-MTRL-2,intro-MTRL-3}. Yet coding tasks have different optimization requirements, with syntactically strict tasks demanding conservative updates for executable correctness and semantically flexible tasks tolerating greater language variation. A uniform strategy therefore over-constrains some tasks while leaving others insufficiently controlled.

\end{itemize}


Inspired by how skilled learners allocate more effort to less-mastered topics and leverage connections between subjects, we introduce \textit{task utility}, a signal capturing both a task's learning potential and its cross-task synergy, and propose \textbf{\tool}, a multi-t\textbf{AS}k code reinforcement learning framework via u\textbf{T}ility-driven co\textbf{OR}dination that resolves both challenges through two coordinated modules.
\textit{\moduleone} uses task utility to hierarchically allocate training budget across tasks and prioritize informative prompts within each task, addressing \textit{what to learn}.
\textit{\moduletwo} uses the same signal to calibrate per-task KL regularization, dynamically adapting policy optimization to each task's training state, addressing \textit{how to learn}.
                                  
We evaluate {\tool} on four coding tasks spanning the software development workflow: code I/O prediction (understanding), code generation (implementation), unit test generation (testing), and commit message generation (documentation), using Qwen2.5-Coder-7B~\cite{Qwen2.5-Coder} and Qwen3-8B~\cite{Qwen3} as base models. Experiments demonstrate that the single {\tool} model surpasses the best task-specific RL specialist by 9.0\%--9.5\% in average performance across all tasks.
{\tool} further surpasses all MTRL baselines by 7.5\%--19.2\% across models and baselines, demonstrating the effectiveness of utility-driven coordination across heterogeneous coding tasks.

The main contributions of our paper are summarized as follows: 




    






\begin{itemize}[leftmargin=*]

    \item To the best of our knowledge, we are the first to study multi-task RL for code LLMs, formulating it around two coupled decisions, what to learn and how to learn, and proposing task utility as a unified signal governing both.

    \item We propose \tool, a coordination framework with hierarchical task allocation and prompt prioritization for data scheduling, and task-adaptive KL regularization for policy optimization.

    \item We empirically validate {\tool} across multiple base LLMs and heterogeneous coding tasks, showing consistent gains over both task-specific RL specialists and MTRL baselines.

\end{itemize}

\section{Methodology}
    \begin{figure*}[t]
    \centering
    \includegraphics[scale=0.355]{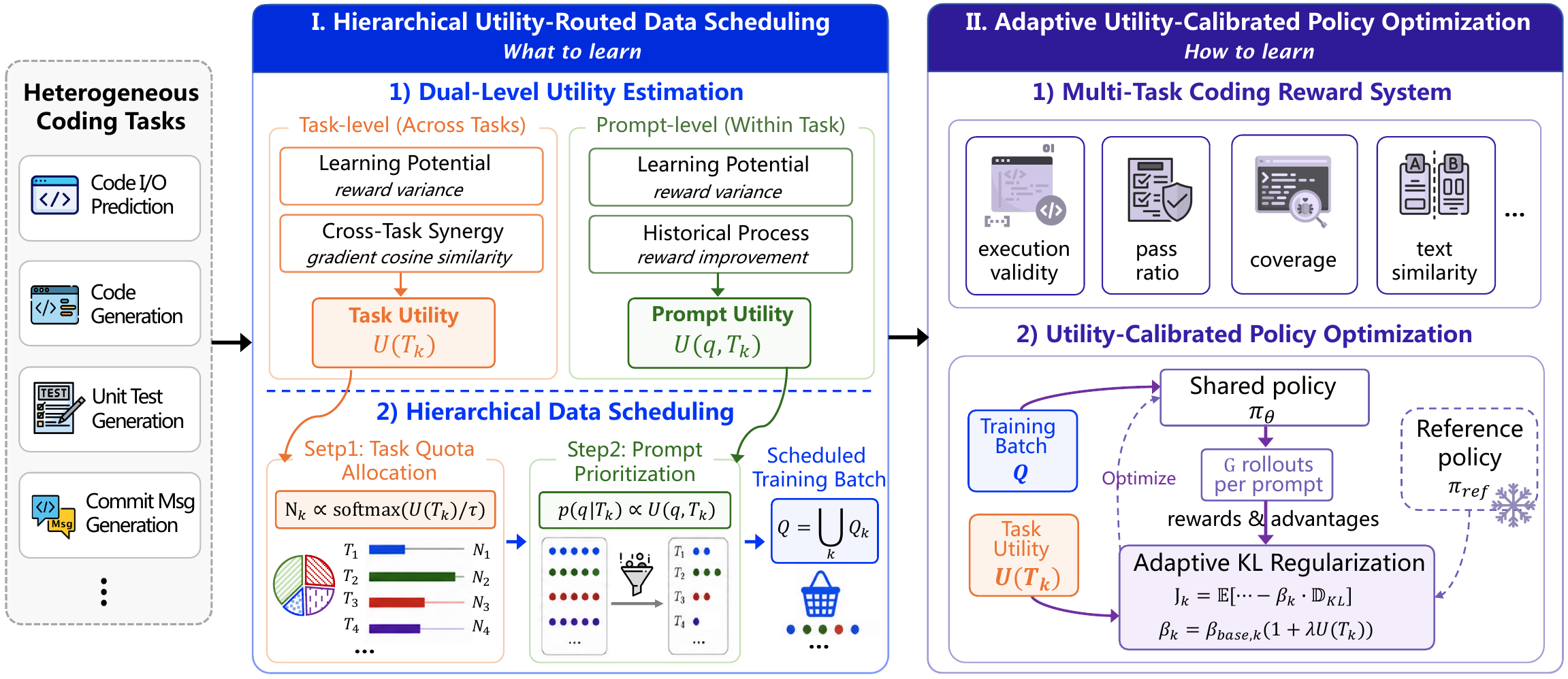}
    \caption{Overview of {{\tool}}. (I) hierarchically schedules training data at task and prompt levels based on estimated task utilities; and (II) updates the shared policy with task-adaptive KL regularization conditioned on task utilities.}
    \label{fig:framework}
\end{figure*}

As illustrated in Figure~\ref{fig:framework}, {\tool} consists of two coupled modules: \textit{(1) \moduleone} hierarchically schedules training data at task and prompt levels based on estimated task utilities;
\textit{(2) \moduletwo} updates the shared policy with task-adaptive KL regularization conditioned on task utilities.

\subsection{Problem Formulation}
\label{subsec:problem_formulation}

We formulate multi-task code post-training as a reinforcement learning problem over $K$ heterogeneous coding tasks $\T=\{T_1,\dots,T_K\}$. Each task $T_k$ provides a prompt dataset $\D_k$ and a reward function $R_k$ for evaluating model rollouts. All tasks share a single policy $\policy$ parameterized by the code LLM, together with a frozen reference policy $\pi_{\text{ref}}$ for regularization.

At each training iteration $t$, we face two coupled decisions that jointly determine how the shared policy improves across heterogeneous tasks:

\paragraph{Decision 1: What to Learn.}
The first decision concerns \textit{data scheduling}, i.e., which tasks and which prompts to select for the next policy update. Under a fixed total batch budget $B$, we allocate task-specific quotas $\{N_k^{(t)}\}_{k=1}^{K}$ and sample prompt batches $\mathcal{Q}_k^{(t)}\subseteq\D_k$ with $|\mathcal{Q}_k^{(t)}|=N_k^{(t)}$ such that
\begin{equation}
\mathcal{Q}^{(t)}=\bigcup_{k=1}^{K}\mathcal{Q}_k^{(t)},
\qquad
\sum_{k=1}^{K}N_k^{(t)}=B.
\end{equation}
This decision determines both \textit{which tasks} appear in the update (via task quotas) and \textit{which prompts} represent each task (via prompt selection).
In multi-task RL, an effective schedule should account for both task-intrinsic signals (e.g., whether a task still has learning potential) and cross-task effects (e.g., whether training on one task benefits others).
Ignoring either aspect leads to suboptimal allocation. Over-sampling saturated tasks wastes computation, while over-sampling conflicting tasks induces negative transfer.

\paragraph{Decision 2: How to Learn.}
The second decision concerns \textit{policy optimization}, i.e., how to update the shared policy on the scheduled data.
We build on GRPO~\cite{grpo} with task-specific KL regularization.
For each prompt $q$, we sample $G$ rollouts $\{o_i\}_{i=1}^G$ from the current policy and maximize:
\begin{equation}
\small
\label{eq:task_grpo_objective}
\begin{aligned}
J_k(\theta;t)
&= \mathbb{E}_{q\sim\mathcal{Q}_k^{(t)},\{o_i\}}
\Bigg[
\frac{1}{G}\sum_{i=1}^{G}
\min\Big(
\rho_{i} A_{i},
\mathrm{clip}(\rho_{i},1-\epsilon,1+\epsilon)A_{i}
\Big) 
- \beta_k(t)\,
\mathbb{D}_{\mathrm{KL}}
\big[\pi_{\theta}(\cdot\mid q)\|\pi_{\text{ref}}(\cdot\mid q)\big]
\Bigg],
\end{aligned}
\end{equation}
where $\rho_{i}=\pi_\theta(o_i|q)/\pi_{\theta_{\text{old}}}(o_i|q)$ is the probability ratio between the current policy and the pre-update policy $\pi_{\theta_{\text{old}}}$, $A_{i}$ is the group-normalized advantage, and $\beta_k(t)$ is a task-specific KL coefficient.

The multi-task objective aggregates task-specific objectives weighted by their batch proportions:
\begin{equation}
\label{eq:mt_objective}
J_{\text{MT}}(\theta;t) = \sum_{k=1}^K \frac{N_k^{(t)}}{B}\,J_k(\theta;t).
\end{equation}

These two decisions are coupled; task quotas $N_k^{(t)}$ and prompt batches $\mathcal{Q}_k^{(t)}$ determine \textit{what to learn}, while KL coefficients $\beta_k(t)$ determine \textit{how to learn}.
In the following, we describe how {\tool} addresses these coupled decisions.
Additional formulation details are provided in Appendix~\ref{sec:appendix_problem_formulation}.

\subsection{\moduleone}
\label{subsec:module1}
This module addresses \textit{what to learn} by using task utility to hierarchically allocate training budget $B$ across tasks and prioritize informative prompts within each task.

\subsubsection{\blockone}
\label{subsubsec:utility_estimation}

Intuitively, a high-utility task should exhibit two key properties: \emph{learning potential}, which reflects actionable optimization signals, and \emph{cross-task synergy}, which captures constructive gradient alignment with other tasks. In other words, such a task should both provide useful learning signals for the current model and facilitate positive transfer across tasks. We therefore define task utility as a signal that jointly captures both properties.

\paragraph{Task Utility.}
\textit{1) Learning Potential.} 
We employ the statistical variance of rollout rewards as an aggregation metric for remaining learning opportunity. High variance indicates a heterogeneous result set where the model yields both successful and failed trajectories. For a sampled query batch $\mathcal{Q}_k \subseteq \mathcal{D}_k$ at step $t$, the potential utility is defined as
\begin{equation}
\mathcal{U}_{\mathsf{pot}}^{(t)}(T_k) = \frac{1}{|\mathcal{Q}_k|} 
\sum_{q \in \mathcal{Q}_k} \mathsf{Var} \left( \{ R_k(q, o_{q,i}) \}_{i=1}^{G} \right)
\end{equation}

\textit{2) Cross-Task Synergy.} High learning potential alone does not guarantee positive cross-task effects.
From a first-order optimization perspective, processing $T_k$ induces a parameter update $\theta \leftarrow \theta + \eta \mathbf{g}_k$, whose effect on task $T_j$ is proportional to $\eta \langle \mathbf{g}_j, \mathbf{g}_k \rangle$, which is positive when the two tasks reinforce each other and negative when they conflict.
This motivates using gradient cosine similarity as a measure of cross-task synergy~\cite{related-PCGrad, GradDrop,CAGrad}.
Let $\mathbf{g}_k$ denote the policy gradient aggregated from task $T_k$; the synergy score is given by
\begin{equation}
\mathcal{U}_{\mathsf{syn}}^{(t)}(T_k) = \frac{1}{|\mathcal{T}|-1} 
\sum_{T_j \in \mathcal{T} \setminus \{T_k\}} \mathsf{sim}(\mathbf{g}_k, \mathbf{g}_j)
\end{equation}


Note that computing full-parameter gradient similarity is infeasible for LLM-scale models; we compress each task gradient by summing over all parameter dimensions except the last within each layer, then concatenating the resulting row vectors across layers into a compact vector per task. The compressed vectors serve as an efficient proxy for measuring task relationships rather than exactly preserving gradient directions: the resulting similarities agree in sign with those computed from full gradients in 92.3\% of cases (Spearman correlation 0.91). Details, validation, and scalability analysis are in Appendix~\ref{sec:appendix_gradient_compression}.


\paragraph{Prompt Utility.}

Within a task, not all prompts are equally informative. A prompt the model already solves provides no useful gradient signal; a prompt the model never solves contributes little beyond reinforcing failures. The most informative prompts are those where the model is actively improving and still uncertain about the correct output. We therefore define prompt-level utility as a combination of current learning potential and historical progress.
Letting $\mu_R^{(t)}(q)$ denote the mean reward over $G$ rollouts at step $t$, we define
\begin{align}
\mathcal{U}_{\mathsf{prog}}^{(t)}(q, T_k) &= \mu_R^{(t)}(q) - \mu_R^{(t-1)}(q) \\
\mathcal{U}_{\mathsf{pot}}^{(t)}(q, T_k) &= \mathsf{Var}\left( \{ R_k(q, o_{q,i}) \}_{i=1}^{G} \right)
\end{align}
where $t-1$ denotes the last earlier step at which $q$ was sampled.


\subsubsection{\blocktwo}
\label{subsubsec:data_schedule}

Equipped with the utility metrics, we formulate batch construction as a hierarchical allocation procedure, first determining the training budget quota per task, and subsequently prioritizing prompts to fulfill each quota.

\paragraph{Task Quota Allocation.}
To allocate the budget $B$ across tasks, we need a single scalar utility per task. The two raw components, however, are not yet comparable: each training step observes only one mini-batch per task, making per-batch statistics noisy, and reward variance and gradient cosine similarity operate on different numerical scales, so one would dominate the combined utility without rescaling. We apply EMA smoothing~\cite{ema} and min-max normalization across tasks to each component before fusion; details are in Appendix~\ref{sec:appendix_scheduler}. The task-level utility is then
\begin{equation}
\mathcal{U}^{(t)}(T_k) = \mathsf{EMA}\!\left( \mathcal{U}_{\mathsf{pot}}^{(t-1)}(T_k) \right) 
+ \mathsf{EMA}\!\left( \mathcal{U}_{\mathsf{syn}}^{(t-1)}(T_k) \right)
\end{equation}
We then partition the global budget $B$ via a temperature-scaled softmax to yield the training quota $N_k^{(t)}$ for task $T_k$,
\begin{equation}
N_k^{(t)} = B \cdot \frac{\exp\left( \mathcal{U}^{(t)}(T_k) / \tau \right)}
{\sum_{T_j \in \mathcal{T}} \exp\left( \mathcal{U}^{(t)}(T_j) / \tau \right)}
\end{equation}
where $\tau > 0$ controls how sharply the budget concentrates on high-utility tasks.

\paragraph{Prompt Prioritization.}
Within each task's quota $N_k^{(t)}$, the same preprocessing is applied at the prompt level,
\begin{equation}
\mathcal{U}^{(t)}(q, T_k) = \mathsf{EMA}\!\left( \mathcal{U}_{\mathsf{pot}}^{(t-1)}(q, T_k) \right) 
+ \mathsf{EMA}\!\left( \mathcal{U}_{\mathsf{prog}}^{(t-1)}(q, T_k) \right)
\end{equation}
Prompts are then selected via a sigmoid-transformed distribution over $\mathcal{D}_k$,
\begin{equation}
\pi^{(t)}(q \mid T_k) = \frac{\sigma\left( \mathcal{U}^{(t)}(q, T_k) \right)}
{\sum_{q' \in \mathcal{D}_k} \sigma\left( \mathcal{U}^{(t)}(q', T_k) \right)}
\end{equation}

and $\mathcal{Q}_k^{(t)}$ is drawn via weighted sampling without replacement. Prompts that have not yet been selected are assigned a weight of 1, the upper bound of $\sigma(\cdot)$, which encourages coverage of $\mathcal{D}_k$ (Appendix~\ref{sec:appendix_scheduler}). These task-specific batches are combined to produce the final heterogeneous training batch $\mathcal{Q}^{(t)} = \bigcup_{k} \mathcal{Q}_k^{(t)}$.
\subsection{\moduletwo}
\label{subsec:module2}

This module addresses \textit{how to learn} by dynamically calibrating the KL regularization coefficient $\beta_k(t)$ for each task based on its current task utility, enabling task-adaptive policy optimization on the heterogeneous training batch produced by the first module.

\subsubsection{\blockthree}
\label{subsubsec:unified_reward}



The reward system defines the interface between the policy and the training environment. {\tool} only requires each task to provide an automatically computable scalar reward. In our experiments, to cover a representative range of software development competencies, we instantiate the reward system across four tasks that span distinct programming paradigms, each grounded in an objective, verifiable reward signal.
\begin{itemize}[leftmargin=*]
    \item \textbf{Code I/O Prediction.} Given a function signature, predict valid input-output pairs. The reward evaluates execution correctness: $\mathcal{R}_{\mathsf{IO}}(o) = \mathbb{I}[\mathsf{Exec}(o) \text{ is valid}] \in \{0, 1\}$.
    \item \textbf{Code Generation.} Given a problem description, synthesize an implementation. The metric is the test case pass rate: $\mathcal{R}_{\mathsf{CG}}(o) = \mathsf{PassRatio}(o)$.
    \item \textbf{Unit Test Generation.} Given a focal function, generate unit tests. The metric calculates line/branch coverage: $\mathcal{R}_{\mathsf{UT}}(o) = \mathsf{Coverage}(o)$.
    \item \textbf{Commit Message Generation.} Given a code diff, generate a commit message. The metric measures textual similarity against ground truth: $\mathcal{R}_{\mathsf{CM}}(o) = \mathsf{TextSim}(o, o_{\mathsf{gt}})$.
\end{itemize}
To ensure structural validity, the final reward objective fuses the task-specific 
metric with a format compliance constraint:
\begin{equation}
\mathcal{R}(o) = w_{T_k} \cdot \mathcal{R}_{T_k}(o) + 
w_{\mathsf{fmt}} \cdot \mathcal{R}_{\mathsf{fmt}}(o)
\end{equation}
where $T_k \in$$ \{\mathsf{IO}, \mathsf{CG},\mathsf{UT},\mathsf{CM}\}$, and $\mathcal{R}_{\mathsf{fmt}} \in \{0,1\}$ acts as a strict boolean filter for output parsing. 
Implementation details of these reward functions, including the concrete weights and format penalty used in our experiments, are provided in Appendix~\ref{sec:appendix_reward}, and prompt templates for all tasks are in Appendix~\ref{sec:appendix_prompts}.

\subsubsection{\blockfour}
\label{subsubsec:dynamic_kl_grpo}

Coding tasks differ in how tightly their outputs are constrained: some must produce complete executable programs, whose syntactic validity is easily broken by large policy shifts, whereas others produce short answers or natural language, for which broader exploration is less risky. As a result, the uniform KL coefficient used by existing methods over-constrains some tasks while under-constraining others, and the appropriate constraint also changes as a task's utility evolves.
We therefore decompose each task's KL coefficient into a task-specific base and a dynamic multiplier conditioned on task utility,
\begin{equation}
\beta_k^{(t)} = \beta_{\mathsf{base},k} \cdot \left(1 + \lambda \cdot \mathcal{U}^{(t)}(T_k)\right)
\end{equation}
where $\beta_{\mathsf{base},k}$ is a task-specific baseline coefficient set according to the task's output constraints and $\lambda$ scales how strongly utility modulates it. The sampling quota and the KL coefficient play complementary roles. A high-utility task receives a larger quota so that training makes greater use of it, and a larger $\beta_k^{(t)}$ so that the shared policy does not over-specialize toward it. Low utility may instead reflect saturation, persistent failure, or task conflict; in each case, the smaller quota already limits the task's contribution, so a smaller $\beta_k^{(t)}$ avoids further suppressing the useful signal its remaining samples may still carry.

Overall, {\tool} operates as an end-to-end coordination framework, alternating between hierarchical 
task allocation and prompt prioritization (data scheduling) and task-adaptive policy optimization. 
The complete algorithm pseudocode is presented in Appendix~\ref{sec:appendix_detailed_algorithm}.

\section{Experiments}
    
\subsection{Experimental Setup}

\textbf{Training Datasets.}
We construct training datasets for four heterogeneous coding tasks: 2,088 samples for \textit{Code I/O Prediction} from LeetCode~\cite{exp-data-LeetCodeDataset}; 2,082 samples for \textit{Code Generation} from LiveCodeBench~\cite{exp-data-lcb} and Codeforces~\cite{exp-data-Codeforces}; 1,148 samples for \textit{Unit Test Generation} from real-world open-source Java repositories with executable validation environments; and 2,446 samples for \textit{Commit Message Generation} from the MCMD dataset~\cite{exp-data-MCMD}. Dataset construction details are provided in Appendix~\ref{sec:appendix_dataset}.

\textbf{Evaluation Benchmarks and Metrics.} We evaluate each task on representative and widely-used benchmarks with task-specific metrics.
For \textit{Code I/O Prediction}, we report input and output accuracy on CRUXEval~\cite{exp-benchmark-crux}.
For \textit{Code Generation}, we measure pass@1 and pass@2 on Aider-Polyglot-Python~\cite{exp-benchmark-Aider}.
For \textit{Unit Test Generation}, we evaluate line coverage, branch coverage, and compilation rate on Defects4J 2.0~\cite{exp-benchmark-defect4j}.
For \textit{Commit Message Generation}, we report BLEU~\cite{exp-metric-BLEU}, ROUGE~\cite{exp-metric-rouge}, and METEOR~\cite{exp-metric-ETEOR} on MCMDEval+~\cite{exp-data-MCMD}. Detailed descriptions of each benchmark and metric are provided in Appendix~\ref{sec:appendix_benchmarks}.

\textbf{Baselines.}
We include \textit{(1) Single-Task Learning (STL)}~\cite{intro-RL-code-4, intro-RL-code-5}: for each task, we train an independent model using GRPO~\cite{grpo} on that task's data only, representing task-specific specialists.

We further compare against three multi-task baselines spanning the main paradigms in LLM-based MTRL, following prior works~\cite{intro-MTRL-2,intro-MTRL-3}.
\textit{(2) Joint Learning}, following Wu et al.~\cite{intro-MTRL-2}: all tasks are uniformly mixed within each training batch and trained with a shared GRPO objective.
\textit{(3) Curriculum Learning}, in the spirit of OMNI-THINKER~\cite{intro-MTRL-3}: tasks are trained sequentially in a fixed order (Code I/O $\to$ Code Gen $\to$ Unit Test $\to$ Commit Message), which follows the software development workflow in Figure~\ref{fig:development} (understanding, implementation, testing, documentation) rather than task difficulty.
\textit{(4) Model Merging}: task-specific GRPO specialists are trained separately, then merged post-hoc via TIES-Merging~\cite{TIES}. All experiments are conducted on Qwen2.5-Coder-7B~\cite{Qwen2.5-Coder} and Qwen3-8B~\cite{Qwen3}. Baseline configurations are detailed in Appendix~\ref{sec:appendix_baselines}.


\textbf{Implementation Details.} 
We use the MindSpeed~\cite{exp-imp-mindspeed} pipeline for RL fine-tuning, with a global batch size of 128 and 8 rollouts per sample. All experiments are trained for 400 steps on 32 Ascend 910B-B3 NPUs. Full hyperparameter settings are provided in Appendix~\ref{sec:appendix_hyperparameters}.

\subsection{Main Results}

Table~\ref{tab:main_results} reports the performance of {\tool} and all baselines across four coding tasks on two base models.

\begin{table*}[t]
    \centering
    \caption{Performance comparison of {\tool} and baselines across four coding tasks. The abbreviation of ``STL'' denotes Single-Task Learning. Bolded values mark the best per column; underscored values mark the second best.}
    \label{tab:main_results}
    \resizebox{\textwidth}{!}{
    \begin{tabular}{l@{\hskip 8pt}cc@{\hskip 8pt}cc@{\hskip 8pt}ccc@{\hskip 8pt}ccc}
    \toprule
    \multirow{3}{*}{\textbf{Method}} & \multicolumn{2}{c}{\textbf{CRUXEval}} & \multicolumn{2}{c}{\textbf{Aider-Polyglot}} & \multicolumn{3}{c}{\textbf{Defects4J}} & \multicolumn{3}{c}{\textbf{MCMDEval+}} \\
    & \multicolumn{2}{c}{\textit{Code I/O Prediction}} & \multicolumn{2}{c}{\textit{Code Generation}} & \multicolumn{3}{c}{\textit{Unit Test Generation}} & \multicolumn{3}{c}{\textit{Commit Message Gen.}} \\
    \cmidrule(lr){2-3} \cmidrule(lr){4-5} \cmidrule(lr){6-8} \cmidrule(lr){9-11}
    & Input & Output & Pass@1 & Pass@2 & Line & Branch & Compile & BLEU & ROUGE & METEOR \\
    & Acc. & Acc. & & & Cov. & Cov. & Rate & & & \\
    \midrule
    \rowcolor{gray!10}
    \multicolumn{1}{l}{\textit{\textbf{Qwen2.5-Coder-7B}}} & \textit{47.8} & \textit{41.0} & \textit{24.4} & \textit{26.7} & \textit{24.4} & \textit{20.5} & \textit{49.3} & \textit{23.8} & \textit{15.5} & \textit{12.9} \\
    \midrule
    STL (Code I/O) & \underline{54.4} & 53.5 & 31.1 & 38.5 & 28.4 & 23.8 & 49.9 & 29.4 & 17.0 & 15.7 \\
    STL (Code Gen) & 51.8 & 52.2 & \underline{40.7} & \underline{44.4} & 28.1 & 23.5 & 51.0 & 28.2 & 17.2 & \underline{17.4} \\
    STL (Unit Test) & 48.6 & 50.4 & 36.3 & 40.7 & \underline{30.8} & \underline{24.6} & \underline{53.0} & 26.0 & 15.7 & 15.6 \\
    STL (Commit Msg) & 52.4 & 50.8 & 19.3 & 27.4 & 29.2 & 23.2 & 50.0 & \textbf{41.2} & \underline{21.4} & 17.3 \\
    \midrule
    Joint Learning & 51.4 & \underline{54.0} & 32.6 & 43.0 & 25.9 & 21.0 & 51.1 & 30.7 & 17.1 & 16.0 \\
    Curriculum Learning & 51.1 & 52.5 & 28.1 & 35.6 & 27.8 & 23.1 & 48.4 & 30.5 & 17.3 & 16.6 \\
    Model Merging & \textbf{54.7} & 53.3 & 23.7 & 31.1 & 27.2 & 22.4 & 46.8 & 30.0 & 17.9 & 17.1 \\
    \midrule
    \rowcolor{blue!18}
    \textbf{\tool}  & 52.2 & \textbf{56.4} & \textbf{41.5} & \textbf{45.2} & \textbf{32.3} & \textbf{25.7} & \textbf{55.7} & \underline{35.2} & \textbf{22.1} & \textbf{20.2} \\
    \midrule[1.5pt]
    \rowcolor{gray!10}
    \multicolumn{1}{l}{\textit{\textbf{Qwen3-8B}}} & \textit{53.3} & \textit{56.3} & \textit{38.5} & \textit{43.0} & \textit{30.9} & \textit{26.9} & \textit{50.2} & \textit{33.8} & \textit{16.6} & \textit{14.8} \\
    \midrule
    STL (Code I/O) & \underline{61.0} & \underline{63.1} & 43.0 & 48.9 & 32.4 & 28.0 & 53.2 & 35.6 & 17.4 & 15.8 \\
    STL (Code Gen) & 57.8 & 61.3 & \underline{50.4} & \underline{53.3} & 31.8 & 27.6 & 52.8 & 35.0 & 17.6 & 16.7 \\
    STL (Unit Test) & 56.5 & 59.8 & 45.2 & 51.1 & \underline{37.2} & \underline{31.2} & \underline{55.9} & 34.0 & 17.0 & 15.2 \\
    STL (Commit Msg) & 57.0 & 60.3 & 40.7 & 45.9 & 32.0 & 27.4 & 52.0 & \underline{40.8} & \underline{21.4} & \underline{18.6} \\
    \midrule
    Joint Learning & 58.8 & 62.5 & 47.4 & 52.6 & 34.0 & 29.1 & 54.4 & 37.2 & 18.8 & 17.3 \\
    Curriculum Learning & 57.3 & 60.6 & 42.2 & 49.6 & 33.1 & 28.5 & 52.9 & 36.6 & 18.4 & 16.8 \\
    Model Merging & 59.5 & 61.5 & 40.0 & 46.7 & 32.6 & 28.2 & 50.7 & 36.2 & 19.1 & 17.6 \\
    \midrule
    \rowcolor{blue!18}
    \textbf{\tool}  & \textbf{61.5} & \textbf{65.1} & \textbf{51.1} & \textbf{54.8} & \textbf{38.1} & \textbf{32.0} & \textbf{56.6} & \textbf{41.2} & \textbf{22.5} & \textbf{20.0} \\
    \bottomrule
    \end{tabular}
    }
    \end{table*}

\textbf{Comparison with Single-Task Learning.}
{\tool} achieves strong multi-task performance that rivals or exceeds task-specific specialists.
For Qwen2.5-Coder-7B, {\tool} surpasses the best single-task model in 8 out of 10 metrics, with an average score of 38.65.
This demonstrates that a single model can handle multiple tasks without significant trade-offs.
Similar trends hold for Qwen3-8B, where {\tool} achieves the best results across all 10 metrics (44.29 vs.\ 40.43). It even exceeds using each task’s own specialist (38.13 and 43.29 on average), which requires four separate models.

\textbf{Comparison with Multi-Task Baselines.}
{\tool} greatly outperforms all multi-task baselines across both models.
On Qwen2.5-Coder-7B, it outperforms Joint Learning (34.28), Curriculum Learning (33.10), and Model Merging (32.42) by 12.8\%, 16.8\%, and 19.2\%, respectively.
Joint Learning suffers from severe task interference (its code generation pass@1 is 21.4\% lower than that of {\tool}); Curriculum Learning is limited by its fixed task ordering; and Model Merging shows that post-hoc integration of specialists is inferior to joint optimization.

\textbf{Generalization Across Model Architectures.}
{\tool} generalizes consistently across model families and scales.
On Qwen2.5-Coder-7B it scores 38.65 (+9.0\% over the best specialist, +12.8\% over Joint Learning), and on Qwen3-8B it reaches 44.29 (+9.5\% and +7.5\%, respectively), confirming that utility-driven scheduling and task-adaptive policy optimization transfer consistently across different model capacities.

\subsection{Ablation Studies}

Table~\ref{tab:ablation} reports ablations of the two key modules of {\tool}: data schedule and policy optimization.

\begin{table*}[t]
    \centering
    \caption{Ablation results of {\tool}.}
    \label{tab:ablation}
    \resizebox{\textwidth}{!}{
    \begin{tabular}{l@{\hskip 8pt}cc@{\hskip 8pt}cc@{\hskip 8pt}ccc@{\hskip 8pt}ccc}
    \toprule
    \multirow{3}{*}{\textbf{Method}} & \multicolumn{2}{c}{\textbf{CRUXEval}} & \multicolumn{2}{c}{\textbf{Aider-Polyglot}} & \multicolumn{3}{c}{\textbf{Defects4J}} & \multicolumn{3}{c}{\textbf{MCMDEval+}} \\
    & \multicolumn{2}{c}{\textit{Code I/O Prediction}} & \multicolumn{2}{c}{\textit{Code Generation}} & \multicolumn{3}{c}{\textit{Unit Test Generation}} & \multicolumn{3}{c}{\textit{Commit Message Gen.}} \\
    \cmidrule(lr){2-3} \cmidrule(lr){4-5} \cmidrule(lr){6-8} \cmidrule(lr){9-11}
    & Input & Output & Pass@1 & Pass@2 & Line & Branch & Compile & BLEU & ROUGE & METEOR \\
    & Acc. & Acc. & & & Cov. & Cov. & Rate & & & \\
    \midrule
    \multicolumn{11}{l}{\cellcolor{gray!10}\textit{\textbf{Qwen2.5-Coder-7B}}} \\
    \midrule
    \rowcolor{blue!18}
    \textbf{\tool}  & \textbf{52.2} & \textbf{56.4} & \textbf{41.5} & \textbf{45.2} & \textbf{32.3} & \textbf{25.7} & \textbf{55.7} & \textbf{35.2} & \textbf{22.1} & \textbf{20.2} \\
    \midrule
    \multicolumn{11}{l}{\quad $\bullet$ \textit{Data Schedule}} \\
    \quad\quad w/o Task Allocation & 55.8 & 55.2 & 34.8 & 41.5 & 28.1 & 23.5 & 51.1 & 33.8 & 19.4 & 17.8 \\
    \quad\quad w/o Prompt Prioritization & 53.6 & 56.7 & 34.8 & 40.7 & 27.8 & 24.1 & 51.5 & 31.9 & 17.9 & 16.7 \\
    \quad\quad w/o Task \& Prompt (Random) & 53.3 & 51.2 & 31.9 & 41.5 & 27.5 & 22.2 & 50.0 & 31.4 & 17.8 & 16.6 \\
    \hdashline
    \multicolumn{11}{l}{\quad $\bullet$ \textit{Policy Optimization}} \\
    \quad\quad w/o Dynamic Regularization & 52.9 & 51.9 & 35.6 & 43.7 & 30.6 & 26.2 & 51.2 & 32.9 & 18.6 & 16.7 \\
    \quad\quad w/o Task-Specific Regularization & 53.6 & 50.7 & 34.1 & 38.5 & 30.6 & 25.5 & 49.9 & 34.3 & 18.1 & 16.3 \\
    \midrule[1.5pt]
    \multicolumn{11}{l}{\cellcolor{gray!10}\textit{\textbf{Qwen3-8B}}} \\
    \midrule
    \rowcolor{blue!18}
    \textbf{\tool} & \textbf{61.5} & \textbf{65.1} & \textbf{51.1} & \textbf{54.8} & \textbf{38.1} & \textbf{32.0} & \textbf{56.6} & \textbf{41.2} & \textbf{22.5} & \textbf{20.0} \\
    \midrule
    \multicolumn{11}{l}{\quad $\bullet$ \textit{Data Schedule}} \\
    \quad\quad w/o Task Allocation & 61.0 & 63.9 & 48.1 & 53.3 & 35.1 & 29.6 & 54.9 & 38.4 & 19.6 & 17.8 \\
    \quad\quad w/o Prompt Prioritization & 60.6 & 64.0 & 49.6 & 53.3 & 36.0 & 30.3 & 55.2 & 38.0 & 19.4 & 17.6 \\
    \quad\quad w/o Task \& Prompt (Random) & 59.1 & 62.6 & 47.4 & 52.6 & 34.3 & 29.2 & 54.4 & 37.3 & 18.9 & 17.3 \\
    \hdashline
    \multicolumn{11}{l}{\quad $\bullet$ \textit{Policy Optimization}} \\
    \quad\quad w/o Dynamic Regularization & 60.3 & 63.5 & 48.9 & 53.3 & 36.8 & 31.0 & 55.4 & 38.8 & 19.9 & 18.0 \\
    \quad\quad w/o Task-Specific Regularization & 60.4 & 63.1 & 48.1 & 52.6 & 36.5 & 30.8 & 54.8 & 39.2 & 20.1 & 18.2 \\
    \bottomrule
    \end{tabular}
    }
    \end{table*}

\textbf{Data Schedule Module.}
We ablate the hierarchical data schedule by removing task allocation, prompt prioritization, or both.

$\blacktriangleright$ \textit{Task Allocation.}
Replacing utility-driven task allocation with uniform sampling causes notable degradation on Qwen2.5-Coder-7B: code generation pass@1 declines by 16.1\% and unit test line coverage by 13.0\%.
Equal resource allocation ignores dynamic learning states, leading to ineffective training and negative transfer.

$\blacktriangleright$ \textit{Prompt Prioritization.}
Replacing utility-based prompt selection with random sampling within each task yields similar declines (pass@1 $-$16.1\%, line coverage $-$13.9\%).
Prioritizing high-utility prompts concentrates training on informative samples, accelerating convergence and reducing wasted computation.

$\blacktriangleright$ \textit{Complete Random Sampling.}
Removing both components produces the worst results (pass@1 $-$23.1\%, line coverage $-$14.9\%), on par with the Joint Learning baseline, confirming that hierarchical utility-driven scheduling is essential for coordinating heterogeneous tasks.

\textbf{Policy Optimization Module.}
We ablate dynamic and task-specific KL regularization.

$\blacktriangleright$ \textit{Dynamic Regularization.}
Fixing KL coefficients throughout training (no utility-based adaptation) degrades code generation pass@1 by 14.2\% and code I/O output accuracy by 8.0\% on Qwen2.5-Coder-7B.
Fixed regularization cannot accommodate evolving task interactions: over-regularization limits synergistic updates while under-regularization enables negative transfer.

$\blacktriangleright$ \textit{Task-Specific Regularization.}
Replacing task-specific KL coefficients with a single uniform value causes code generation pass@2 to decline by 14.8\% and unit test compilation rate to drop by 10.4\% on Qwen2.5-Coder-7B.
Tasks are fundamentally heterogeneous: text-generation tasks require lower regularization for diverse reasoning, while syntactic tasks like unit test generation need stronger constraints for correctness, and a uniform coefficient cannot accommodate both.

\subsection{Further Analysis}

We analyze key training dynamics and hyperparameter sensitivity of {\tool} on Qwen2.5-Coder-7B.

\begin{table*}[t]
    \centering
    \caption{Sensitivity analysis of (a) softmax temperature $\tau$ in task allocation and (b) KL adjustment rate $\lambda$ in dynamic regularization, evaluated on Qwen2.5-Coder-7B.}
    \label{tab:sensitivity}
    \resizebox{\textwidth}{!}{
    \begin{tabular}{l@{\hskip 10pt}cc@{\hskip 8pt}cc@{\hskip 8pt}ccc@{\hskip 8pt}ccc@{\hskip 8pt}c}
    \toprule
    \multirow{3}{*}{\textbf{Hyperparameter}} &
      \multicolumn{2}{c}{\textbf{CRUXEval}} &
      \multicolumn{2}{c}{\textbf{Aider-Polyglot}} &
      \multicolumn{3}{c}{\textbf{Defects4J}} &
      \multicolumn{3}{c}{\textbf{MCMDEval+}} &
      \multirow{3}{*}{\textbf{Avg}} \\
    & \multicolumn{2}{c}{\textit{Code I/O Prediction}} &
      \multicolumn{2}{c}{\textit{Code Generation}} &
      \multicolumn{3}{c}{\textit{Unit Test Generation}} &
      \multicolumn{3}{c}{\textit{Commit Message Gen.}} & \\
    \cmidrule(lr){2-3}\cmidrule(lr){4-5}\cmidrule(lr){6-8}\cmidrule(lr){9-11}
    & Input Acc. & Output Acc. & Pass@1 & Pass@2 & Line Cov. & Branch Cov. & Compile Rate & BLEU & ROUGE & METEOR & \\
    \midrule
    \multicolumn{12}{l}{\cellcolor{gray!10}\quad\textit{(a) Softmax Temperature $\tau$ — Task Allocation Module}} \\
    \midrule
    $\tau = 0.5$ & 50.4 & 53.8 & \textbf{42.2} & \textbf{45.9} & \textbf{33.5} & \textbf{26.2} & \textbf{56.4} & 33.4 & 20.4 & 18.4 & 38.06 \\
    \rowcolor{blue!18}
    $\tau = \mathbf{1.0}$& 52.2 & \textbf{56.4} & 41.5 & 45.2 & 32.3 & 25.7 & 55.7 & \textbf{35.2} & \textbf{22.1} & \textbf{20.2} & \textbf{38.65} \\
    $\tau = 2.0$ & \textbf{53.8} & 55.3 & 37.8 & 42.2 & 29.8 & 23.8 & 52.8 & 34.0 & 20.6 & 18.8 & 36.89 \\
    \midrule[1.5pt]
    \multicolumn{12}{l}{\cellcolor{gray!10}\quad\textit{(b) KL Adjustment Rate $\lambda$ — Policy Optimization Module}} \\
    \midrule
    $\lambda = 0.1$ & 52.5 & 54.3 & 38.5 & 43.7 & 31.2 & 25.3 & 53.5 & 34.0 & 20.3 & 18.5 & 37.18 \\
    \rowcolor{blue!18}
    $\lambda = \mathbf{0.2}$& 52.2 & \textbf{56.4} & \textbf{41.5} & \textbf{45.2} & \textbf{32.3} & \textbf{25.7} & \textbf{55.7} & \textbf{35.2} & \textbf{22.1} & \textbf{20.2} & \textbf{38.65} \\
    $\lambda = 0.4$ & 52.4 & 55.6 & 40.0 & 44.4 & 31.8 & 25.2 & 54.4 & 34.4 & 21.0 & 19.2 & 37.84 \\
    $\lambda = 0.8$ & \textbf{53.0} & 52.8 & 37.0 & 42.2 & 30.4 & 24.6 & 52.0 & 33.4 & 19.0 & 17.4 & 36.18 \\
    \bottomrule
    \end{tabular}
    }
\end{table*}

\begin{wrapfigure}{r}{0.38\linewidth}
    \vspace{-0.8em}
    \centering
    \begin{subfigure}[t]{\linewidth}
        \centering
        \includegraphics[width=\linewidth]{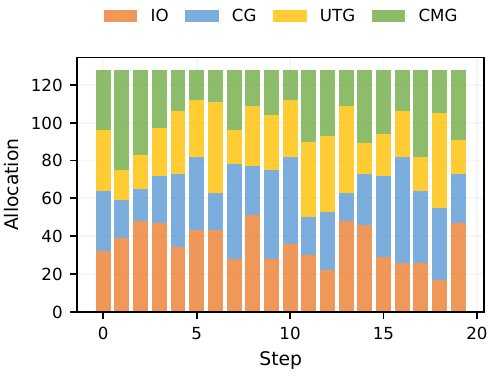}
        \caption{Task allocation over training.}
        \label{fig:further_allocation}
    \end{subfigure}
    \vspace{0.2em}
    \begin{subfigure}[t]{\linewidth}
        \centering
        \includegraphics[width=\linewidth]{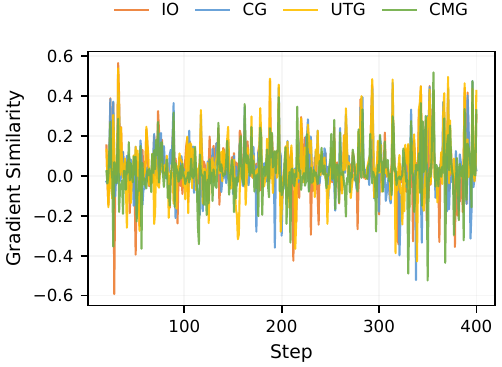}
        \caption{Cross-task gradient similarity.}
        \label{fig:further_gradient}
    \end{subfigure}
    \caption{Training dynamics of {\tool} on Qwen2.5-Coder-7B.}
    \label{fig:further_analysis}
    \vspace{-1em}
\end{wrapfigure}

\textbf{Impact of $\tau$.}
As shown in Table~\ref{tab:sensitivity}(a), a moderate $\tau$ yields the best performance; deviations in either direction lead to degradation.
Too small a $\tau$ causes the scheduler to over-concentrate budget on high-utility tasks, depriving lower-utility tasks of sufficient training signal.
Too large a $\tau$ flattens the allocation toward uniform sampling, diminishing the effect of utility-driven scheduling.

\textbf{Impact of $\lambda$.}
As shown in Table~\ref{tab:sensitivity}(b), performance is relatively consistent across moderate values of $\lambda$ but degrades at both extremes.
When $\lambda$ is too small, the utility-based adjustment becomes negligible, effectively reducing to a static KL scheme that cannot adapt to each task's training dynamics.
When $\lambda$ is too large, high-utility tasks receive overly tight regularization that limits policy updates even where room for improvement remains.

\textbf{Training Dynamics Analysis.}
As shown in Figure~\ref{fig:further_allocation}, task allocation evolves continuously throughout training, with each task's quota rising or falling in response to its real-time utility, a pattern that would be invisible to any fixed schedule.
Figure~\ref{fig:further_gradient} reveals why task-specific regularization matters. Unit test generation exhibits persistent gradient conflicts with other tasks (drops $<\!-0.4$), while code I/O and commit message generation remain largely synergistic, underscoring the need to treat tasks differently at the policy level.
Qualitative examples of model outputs across tasks are provided in Appendix~\ref{sec:appendix_examples}.

\section{Related Work}
    \textbf{Multi-Task Learning.}
Multi-task learning~\cite{intro-MTRL-def} trains a shared model on multiple tasks, where gradient interference is a central challenge.
Gradient-manipulation methods mitigate this at the update level; for example, PCGrad~\cite{related-PCGrad} projects conflicting gradients onto orthogonal planes. In multi-task RL, Distral~\cite{intro-MTRL-Tra-4} shares a distilled centroid policy across task workers, using KL regularization to couple separate task-specific policies with a shared one.
In LLM post-training, Wu et al.~\cite{intro-MTRL-2} show that naive task mixing induces severe gradient imbalance and that sampling tasks in proportion to gradient magnitude brings no consistent gain; OMNI-THINKER~\cite{intro-MTRL-3} addresses cross-task forgetting via backward-transfer-guided curriculum scheduling with a task order fixed before training.
Unlike these works, {\tool} estimates task utility online and uses it to decide, within a single shared policy, both the composition of each batch and the regularization strength of each task.
A detailed discussion is in Appendix~\ref{sec:extended_related_work}.

\textbf{RL Post-Training for Code LLMs.}
RL with verifiable rewards has emerged as the dominant post-training paradigm for code LLMs~\cite{grpo}.
For example, RLEF~\cite{intro-RL-code-4} trains code models on multi-turn execution feedback, CodeRL+~\cite{intro-RL-code-2} integrates execution semantics alignment into the RL objective, and Lee et al.~\cite{intro-RL-code-5} apply adversarial RL for unit test generation. All of these approaches train task-specific specialists that degrade sharply outside their training domain. {\tool} addresses this by jointly post-training a single model across various heterogeneous coding tasks under a unified multi-task RL framework. A detailed discussion is in Appendix~\ref{sec:extended_related_work}.

\section{Conclusion}
    In this paper, we propose \tool, a utility-driven coordination framework that addresses both through a shared signal, \textit{task utility}. \textit{\moduleone} uses task utility to hierarchically allocate training budget across tasks and prioritize informative prompts within each task; \textit{\moduletwo} uses the same signal to calibrate per-task KL regularization, enabling task-adaptive policy optimization across heterogeneous tasks. Experiments across four representative coding tasks and two base LLMs show that {\tool} consistently surpasses both task-specific RL specialists and multi-task baselines, validating the effectiveness of utility-driven coordination for multi-task code RL with a single model. Limitations and future work are discussed in Appendix~\ref{sec:limit_future}.





{\small
\bibliography{ref}

@article{intro-RL-code-1,
  author       = {Daya Guo and
                  Dejian Yang and
                  Haowei Zhang and
                  Junxiao Song and
                  Peiyi Wang and
                  Qihao Zhu and
                  Runxin Xu and
                  Ruoyu Zhang and
                  Shirong Ma and
                  Xiao Bi and
                  Xiaokang Zhang and
                  Xingkai Yu and
                  Yu Wu and
                  Z. F. Wu and
                  Zhibin Gou and
                  Zhihong Shao and
                  Zhuoshu Li and
                  Ziyi Gao and
                  Aixin Liu and
                  Bing Xue and
                  Bingxuan Wang and
                  Bochao Wu and
                  Bei Feng and
                  Chengda Lu and
                  Chenggang Zhao and
                  Chengqi Deng and
                  Chong Ruan and
                  Damai Dai and
                  Deli Chen and
                  Dongjie Ji and
                  Erhang Li and
                  Fangyun Lin and
                  Fucong Dai and
                  Fuli Luo and
                  Guangbo Hao and
                  Guanting Chen, et al},
  title        = {DeepSeek-R1 incentivizes reasoning in LLMs through reinforcement learning},
  journal      = {Nat.},
  volume       = {645},
  number       = {8081},
  pages        = {633--638},
  year         = {2025}
}

@article{intro-RL-code-2,
  author       = {Xue Jiang and
                  Yihong Dong and
                  Mengyang Liu and
                  Hongyi Deng and
                  Tian Wang and
                  Yongding Tao and
                  Rongyu Cao and
                  Binhua Li and
                  Zhi Jin and
                  Wenpin Jiao and
                  Fei Huang and
                  Yongbin Li and
                  Ge Li},
  title        = {CodeRL+: Improving Code Generation via Reinforcement with Execution
                  Semantics Alignment},
  journal      = {CoRR},
  volume       = {abs/2510.18471},
  year         = {2025}
}

@article{intro-RL-code-3,
  author       = {Kimi Team and
                  Angang Du and
                  Bofei Gao and
                  Bowei Xing and
                  Changjiu Jiang and
                  Cheng Chen and
                  Cheng Li and
                  Chenjun Xiao and
                  Chenzhuang Du and
                  Chonghua Liao and
                  Chuning Tang and
                  Congcong Wang and
                  Dehao Zhang and
                  Enming Yuan and
                  Enzhe Lu and
                  Fengxiang Tang and
                  Flood Sung and
                  Guangda Wei and
                  Guokun Lai and
                  Haiqing Guo and
                  Han Zhu and
                  Hao Ding and
                  Hao Hu and
                  Hao Yang and
                  Hao Zhang and
                  Haotian Yao and
                  Haotian Zhao and
                  Haoyu Lu and
                  Haoze Li and
                  Haozhen Yu, et al},
  title        = {Kimi k1.5: Scaling Reinforcement Learning with LLMs},
  journal      = {CoRR},
  volume       = {abs/2501.12599},
  year         = {2025}
}

@inproceedings{intro-RL-code-4,
  author       = {Jonas Gehring and
                  Kunhao Zheng and
                  Jade Copet and
                  Vegard Mella and
                  Taco Cohen and
                  Gabriel Synnaeve},
  title        = {{RLEF:} Grounding Code LLMs in Execution Feedback with Reinforcement
                  Learning},
  booktitle    = {Forty-second International Conference on Machine Learning, {ICML}
                  2025, Vancouver, BC, Canada, July 13-19, 2025},
  publisher    = {OpenReview.net},
  year         = {2025}
}

@article{intro-RL-code-5,
  author       = {Dongjun Lee and
                  Changho Hwang and
                  Kimin Lee},
  title        = {Learning to Generate Unit Test via Adversarial Reinforcement Learning},
  journal      = {CoRR},
  volume       = {abs/2508.21107},
  year         = {2025}
}

@article{intro-MTRL-2,
  author       = {Runzhe Wu and
                  Ankur Samanta and
                  Ayush Jain and
                  Scott Fujimoto and
                  Jeongyeol Kwon and
                  Ben Kretzu and
                  Youliang Yu and
                  Kaveh Hassani and
                  Boris Vidolov and
                  Yonathan Efroni},
  title        = {Imbalanced Gradients in {RL} Post-Training of Multi-Task LLMs},
  journal      = {CoRR},
  volume       = {abs/2510.19178},
  year         = {2025}
}

@inproceedings{intro-MTRL-3,
  title={Omni-thinker: Scaling cross-domain generalization in llms via multi-task rl with hybrid rewards},
  author={Li, Derek and Zhou, Jiaming and Kazemi, Amirreza and Sun, Qianyi and Ghaddar, Abbas and Ma, Liheng and Luo, Yu and Li, Dong and HAO, Jianye and Zhang, Yingxue},
  booktitle={2nd AI for Math Workshop@ ICML 2025},
  year         = {2025}
}

@article{Qwen2.5-Coder,
  author       = {Binyuan Hui and
                  Jian Yang and
                  Zeyu Cui and
                  Jiaxi Yang and
                  Dayiheng Liu and
                  Lei Zhang and
                  Tianyu Liu and
                  Jiajun Zhang and
                  Bowen Yu and
                  Kai Dang and
                  An Yang and
                  Rui Men and
                  Fei Huang and
                  Xingzhang Ren and
                  Xuancheng Ren and
                  Jingren Zhou and
                  Junyang Lin},
  title        = {Qwen2.5-Coder Technical Report},
  journal      = {CoRR},
  volume       = {abs/2409.12186},
  year         = {2024}
}

@article{Qwen3,
  author       = {An Yang and
                  Anfeng Li and
                  Baosong Yang and
                  Beichen Zhang and
                  Binyuan Hui and
                  Bo Zheng and
                  Bowen Yu and
                  Chang Gao and
                  Chengen Huang and
                  Chenxu Lv and
                  Chujie Zheng and
                  Dayiheng Liu and
                  Fan Zhou and
                  Fei Huang and
                  Feng Hu and
                  Hao Ge and
                  Haoran Wei and
                  Huan Lin and
                  Jialong Tang and
                  Jian Yang and
                  Jianhong Tu and
                  Jianwei Zhang and
                  Jian Yang and
                  Jiaxi Yang and
                  Jingren Zhou and
                  Junyang Lin, et al},
  title        = {Qwen3 Technical Report},
  journal      = {CoRR},
  volume       = {abs/2505.09388},
  year         = {2025}
}

@inproceedings{intro-MTRL-Tra-1,
  author       = {Tung{-}Long Vuong and
                  Do Van Nguyen and
                  Tai{-}Long Nguyen and
                  Cong{-}Minh Bui and
                  Hai{-}Dang Kieu and
                  Viet{-}Cuong Ta and
                  Quoc{-}Long Tran and
                  Thanh Ha Le},
  editor       = {Sarit Kraus},
  title        = {Sharing Experience in Multitask Reinforcement Learning},
  booktitle    = {Proceedings of the Twenty-Eighth International Joint Conference on
                  Artificial Intelligence, {IJCAI} 2019, Macao, China, August 10-16,
                  2019},
  pages        = {3642--3648},
  publisher    = {ijcai.org},
  year         = {2019}
}

@inproceedings{intro-MTRL-Tra-2,
  author       = {Ruihan Yang and
                  Huazhe Xu and
                  Yi Wu and
                  Xiaolong Wang},
  editor       = {Hugo Larochelle and
                  Marc'Aurelio Ranzato and
                  Raia Hadsell and
                  Maria{-}Florina Balcan and
                  Hsuan{-}Tien Lin},
  title        = {Multi-Task Reinforcement Learning with Soft Modularization},
  booktitle    = {Advances in Neural Information Processing Systems 33: Annual Conference
                  on Neural Information Processing Systems 2020, NeurIPS 2020, December
                  6-12, 2020, virtual},
  year         = {2020}
}

@inproceedings{intro-MTRL-Tra-3,
  author       = {Myungsik Cho and
                  Jongeui Park and
                  Suyoung Lee and
                  Youngchul Sung},
  title        = {Hard Tasks First: Multi-Task Reinforcement Learning Through Task Scheduling},
  booktitle    = {Forty-first International Conference on Machine Learning, {ICML} 2024,
                  Vienna, Austria, July 21-27, 2024},
  publisher    = {OpenReview.net},
  year         = {2024}
}

@article{intro-MTRL-def,
  author       = {Rich Caruana},
  title        = {Multitask Learning},
  journal      = {Mach. Learn.},
  volume       = {28},
  number       = {1},
  pages        = {41--75},
  year         = {1997}
}

@inproceedings{intro-MTRL-Tra-4,
  author       = {Yee Whye Teh and
                  Victor Bapst and
                  Wojciech M. Czarnecki and
                  John Quan and
                  James Kirkpatrick and
                  Raia Hadsell and
                  Nicolas Heess and
                  Razvan Pascanu},
  editor       = {Isabelle Guyon and
                  Ulrike von Luxburg and
                  Samy Bengio and
                  Hanna M. Wallach and
                  Rob Fergus and
                  S. V. N. Vishwanathan and
                  Roman Garnett},
  title        = {Distral: Robust multitask reinforcement learning},
  booktitle    = {Advances in Neural Information Processing Systems 30: Annual Conference
                  on Neural Information Processing Systems 2017, December 4-9, 2017,
                  Long Beach, CA, {USA}},
  pages        = {4496--4506},
  year         = {2017}
}

@article{exp-data-Codeforces,
  author       = {Yujia Li and
                  David H. Choi and
                  Junyoung Chung and
                  Nate Kushman and
                  Julian Schrittwieser and
                  R{\'{e}}mi Leblond and
                  Tom Eccles and
                  James Keeling and
                  Felix Gimeno and
                  Agustin Dal Lago and
                  Thomas Hubert and
                  Peter Choy and
                  Cyprien de Masson d'Autume and
                  Igor Babuschkin and
                  Xinyun Chen and
                  Po{-}Sen Huang and
                  Johannes Welbl and
                  Sven Gowal and
                  Alexey Cherepanov and
                  James Molloy and
                  Daniel J. Mankowitz and
                  Esme Sutherland Robson and
                  Pushmeet Kohli and
                  Nando de Freitas and
                  Koray Kavukcuoglu and
                  Oriol Vinyals},
  title        = {Competition-Level Code Generation with AlphaCode},
  journal      = {CoRR},
  volume       = {abs/2203.07814},
  year         = {2022}
}

@inproceedings{exp-data-lcb,
  author       = {Naman Jain and
                  King Han and
                  Alex Gu and
                  Wen{-}Ding Li and
                  Fanjia Yan and
                  Tianjun Zhang and
                  Sida Wang and
                  Armando Solar{-}Lezama and
                  Koushik Sen and
                  Ion Stoica},
  title        = {LiveCodeBench: Holistic and Contamination Free Evaluation of Large
                  Language Models for Code},
  booktitle    = {The Thirteenth International Conference on Learning Representations,
                  {ICLR} 2025, Singapore, April 24-28, 2025},
  publisher    = {OpenReview.net},
  year         = {2025}
}

@article{exp-data-LeetCodeDataset,
  author       = {Yunhui Xia and
                  Wei Shen and
                  Yan Wang and
                  Jason Klein Liu and
                  Huifeng Sun and
                  Siyue Wu and
                  Jian Hu and
                  Xiaolong Xu},
  title        = {LeetCodeDataset: {A} Temporal Dataset for Robust Evaluation and Efficient
                  Training of Code LLMs},
  journal      = {CoRR},
  volume       = {abs/2504.14655},
  year         = {2025}
}

@article{exp-data-MCMD,
  author       = {Pengyu Xue and
                  Linhao Wu and
                  Zhongxing Yu and
                  Zhi Jin and
                  Zhen Yang and
                  Xinyi Li and
                  Zhenyu Yang and
                  Yue Tan},
  title        = {Automated Commit Message Generation With Large Language Models: An
                  Empirical Study and Beyond},
  journal      = {{IEEE} Trans. Software Eng.},
  volume       = {50},
  number       = {12},
  pages        = {3208--3224},
  year         = {2024}
}

@misc{exp-benchmark-Aider,
  author = {Paul Gauthier},
  title  = {Aider LLM Leaderboard},
  year   = {2025},
  url    = {https://aider.chat/docs/leaderboards/}
}

@inproceedings{exp-benchmark-crux,
  author       = {Alex Gu and
                  Baptiste Rozi{\`{e}}re and
                  Hugh James Leather and
                  Armando Solar{-}Lezama and
                  Gabriel Synnaeve and
                  Sida Wang},
  title        = {CRUXEval: {A} Benchmark for Code Reasoning, Understanding and Execution},
  booktitle    = {Forty-first International Conference on Machine Learning, {ICML} 2024,
                  Vienna, Austria, July 21-27, 2024},
  publisher    = {OpenReview.net},
  year         = {2024}
}

@inproceedings{exp-benchmark-defect4j,
  author       = {Ren{\'{e}} Just and
                  Darioush Jalali and
                  Michael D. Ernst},
  editor       = {Corina S. Pasareanu and
                  Darko Marinov},
  title        = {Defects4J: a database of existing faults to enable controlled testing
                  studies for Java programs},
  booktitle    = {International Symposium on Software Testing and Analysis, {ISSTA}
                  '14, San Jose, CA, {USA} - July 21 - 26, 2014},
  pages        = {437--440},
  publisher    = {{ACM}},
  year         = {2014}
}

@inproceedings{exp-metric-BLEU,
  author       = {Kishore Papineni and
                  Salim Roukos and
                  Todd Ward and
                  Wei{-}Jing Zhu},
  title        = {Bleu: a Method for Automatic Evaluation of Machine Translation},
  booktitle    = {Proceedings of the 40th Annual Meeting of the Association for Computational
                  Linguistics, July 6-12, 2002, Philadelphia, PA, {USA}},
  pages        = {311--318},
  publisher    = {{ACL}},
  year         = {2002}
}

@inproceedings{exp-metric-rouge,
  title={Rouge: A package for automatic evaluation of summaries},
  author={Lin, Chin-Yew},
  booktitle={Text summarization branches out},
  pages={74--81},
  year={2004}
}

@inproceedings{exp-metric-ETEOR,
  author       = {Satanjeev Banerjee and
                  Alon Lavie},
  editor       = {Jade Goldstein and
                  Alon Lavie and
                  Chin{-}Yew Lin and
                  Clare R. Voss},
  title        = {{METEOR:} An Automatic Metric for {MT} Evaluation with Improved Correlation
                  with Human Judgments},
  booktitle    = {Proceedings of the Workshop on Intrinsic and Extrinsic Evaluation
                  Measures for Machine Translation and/or Summarization@ACL 2005, Ann
                  Arbor, Michigan, USA, June 29, 2005},
  pages        = {65--72},
  publisher    = {Association for Computational Linguistics},
  year         = {2005}
}

@article{grpo,
  author       = {Zhihong Shao and
                  Peiyi Wang and
                  Qihao Zhu and
                  Runxin Xu and
                  Junxiao Song and
                  Mingchuan Zhang and
                  Y. K. Li and
                  Y. Wu and
                  Daya Guo},
  title        = {DeepSeekMath: Pushing the Limits of Mathematical Reasoning in Open
                  Language Models},
  journal      = {CoRR},
  volume       = {abs/2402.03300},
  year         = {2024}
}

@article{TIES,
  author       = {Han Wu and
                  Yuxuan Yao and
                  Shuqi Liu and
                  Zehua Liu and
                  Xiaojin Fu and
                  Xiongwei Han and
                  Xing Li and
                  Hui{-}Ling Zhen and
                  Tao Zhong and
                  Mingxuan Yuan},
  title        = {Unlocking Efficient Long-to-Short {LLM} Reasoning with Model Merging},
  journal      = {CoRR},
  volume       = {abs/2503.20641},
  year         = {2025}
}

@article{exp-imp-mindspeed,
  author       = {Laingjun Feng and
                  Chenyi Pan and
                  Xinjie Guo and
                  Fei Mei and
                  Benzhe Ning and
                  Jianxiang Zhang and
                  Xinyang Liu and
                  Beirong Zhou and
                  Zeng Shu and
                  Chang Liu and
                  Guang Yang and
                  Zhenyu Han and
                  Jiangben Wang and
                  Bo Wang},
  title        = {MindSpeed {RL:} Distributed Dataflow for Scalable and Efficient {RL}
                  Training on Ascend {NPU} Cluster},
  journal      = {CoRR},
  volume       = {abs/2507.19017},
  year         = {2025}
}

@article{ema,
  title={The exponentially weighted moving average},
  author={Hunter, J Stuart},
  journal={Journal of quality technology},
  volume={18},
  number={4},
  pages={203--210},
  year={1986}
}

@inproceedings{related-PCGrad,
  author       = {Tianhe Yu and
                  Saurabh Kumar and
                  Abhishek Gupta and
                  Sergey Levine and
                  Karol Hausman and
                  Chelsea Finn},
  editor       = {Hugo Larochelle and
                  Marc'Aurelio Ranzato and
                  Raia Hadsell and
                  Maria{-}Florina Balcan and
                  Hsuan{-}Tien Lin},
  title        = {Gradient Surgery for Multi-Task Learning},
  booktitle    = {Advances in Neural Information Processing Systems 33: Annual Conference
                  on Neural Information Processing Systems 2020, NeurIPS 2020, December
                  6-12, 2020, virtual},
  year         = {2020}
}

@inproceedings{metagpt,
  author       = {Sirui Hong and
                  Mingchen Zhuge and
                  Jonathan Chen and
                  Xiawu Zheng and
                  Yuheng Cheng and
                  Jinlin Wang and
                  Ceyao Zhang and
                  Zili Wang and
                  Steven Ka Shing Yau and
                  Zijuan Lin and
                  Liyang Zhou and
                  Chenyu Ran and
                  Lingfeng Xiao and
                  Chenglin Wu and
                  J{\"{u}}rgen Schmidhuber},
  title        = {MetaGPT: Meta Programming for {A} Multi-Agent Collaborative Framework},
  booktitle    = {The Twelfth International Conference on Learning Representations,
                  {ICLR} 2024, Vienna, Austria, May 7-11, 2024},
  publisher    = {OpenReview.net},
  year         = {2024}
}

@article{se,
  author       = {Cuiyun Gao and
                  Xing Hu and
                  Shan Gao and
                  Xin Xia and
                  Zhi Jin},
  title        = {The Current Challenges of Software Engineering in the Era of Large
                  Language Models},
  journal      = {{ACM} Trans. Softw. Eng. Methodol.},
  volume       = {34},
  number       = {5},
  pages        = {127:1--127:30},
  year         = {2025}
}

@article{se2,
  author       = {Kaixin Wang and
                  Tianlin Li and
                  Xiaoyu Zhang and
                  Chong Wang and
                  Weisong Sun and
                  Yang Liu and
                  Bin Shi},
  title        = {Software Development Life Cycle Perspective: {A} Survey of Benchmarks
                  for Code Large Language Models and Agents},
  journal      = {CoRR},
  volume       = {abs/2505.05283},
  year         = {2025}
}

@inproceedings{CAGrad,
  author       = {Bo Liu and
                  Xingchao Liu and
                  Xiaojie Jin and
                  Peter Stone and
                  Qiang Liu},
  editor       = {Marc'Aurelio Ranzato and
                  Alina Beygelzimer and
                  Yann N. Dauphin and
                  Percy Liang and
                  Jennifer Wortman Vaughan},
  title        = {Conflict-Averse Gradient Descent for Multi-task learning},
  booktitle    = {Advances in Neural Information Processing Systems 34: Annual Conference
                  on Neural Information Processing Systems 2021, NeurIPS 2021, December
                  6-14, 2021, virtual},
  pages        = {18878--18890},
  year         = {2021}
}

@inproceedings{GradDrop,
  author       = {Zhao Chen and
                  Jiquan Ngiam and
                  Yanping Huang and
                  Thang Luong and
                  Henrik Kretzschmar and
                  Yuning Chai and
                  Dragomir Anguelov},
  editor       = {Hugo Larochelle and
                  Marc'Aurelio Ranzato and
                  Raia Hadsell and
                  Maria{-}Florina Balcan and
                  Hsuan{-}Tien Lin},
  title        = {Just Pick a Sign: Optimizing Deep Multitask Models with Gradient Sign
                  Dropout},
  booktitle    = {Advances in Neural Information Processing Systems 33: Annual Conference
                  on Neural Information Processing Systems 2020, NeurIPS 2020, December
                  6-12, 2020, virtual},
  year         = {2020}
}

@inproceedings{cmg-llm-judge,
  title={Evaluating generated commit messages with large language models},
  author={Zeng, Qunhong and Zhang, Yuxia and Ma, Zexiong and Jiang, Bo and Sun, Ningyuan and Stol, Klaas-Jan and Mou, Xingyu and Liu, Hui},
  booktitle={Proceedings of the 2026 IEEE/ACM 48th International Conference on Software Engineering},
  pages={362--373},
  year={2026}
}

@article{DBLP:journals/pacmse/ChenYLMG25,
  author       = {Yujia Chen and
                  Yang Ye and
                  Zhongqi Li and
                  Yuchi Ma and
                  Cuiyun Gao},
  title        = {Smaller but Better: Self-Paced Knowledge Distillation for Lightweight
                  yet Effective LCMs},
  journal      = {Proc. {ACM} Softw. Eng.},
  volume       = {2},
  number       = {{FSE}},
  pages        = {3057--3080},
  year         = {2025}
}
\bibliographystyle{unsrtnat}
}

\clearpage
\appendix
\section*{\huge{Appendix}}
\section{Methodology Details}
\label{sec:appendix_methodology}

\subsection{Problem Formulation Details}
\label{sec:appendix_problem_formulation}

\paragraph{Rollout Generation.}
Given prompt $q\sim \D_k$, the policy samples a rollout $o=(o_1,\dots,o_{|o|})$ token-by-token from $\pi_{\theta}(\cdot\mid q)$.
We write $o_{<t}=(o_1,\dots,o_{t-1})$ for the prefix at step $t$.
At training iteration $t$, let $\pi_{\theta_{\text{old}}}$ denote the policy before the update.

\paragraph{Training Batch Construction.}
We draw a per-task prompt batch $\mathcal{Q}_k^{(t)}\subseteq \D_k$ of size $|\mathcal{Q}_k^{(t)}|=N_k^{(t)}$, and form the mixed prompt batch $\mathcal{Q}^{(t)}=\bigcup_{k=1}^K \mathcal{Q}_k^{(t)}$ of total size $B$, with
\begin{equation}
    \sum_{k=1}^K N_k^{(t)} = B.
\end{equation}
For each prompt $q\in \mathcal{Q}_k^{(t)}$, we sample a group of $G$ rollouts from $\pi_{\theta_{\text{old}}}(\cdot\mid q)$ and evaluate rewards with $R_k$.

\subsection{Gradient Compression for Efficient Similarity Computation}
\label{sec:appendix_gradient_compression}

Computing cosine similarity on the full gradient vector ($\sim 10^9$ dimensions for a 7B model) is infeasible at training time.
Our goal is to detect whether task gradients are aligned or in conflict, rather than to recover their exact directions; we therefore use a compressed representation as an efficient proxy for task relationships.
We compress each task gradient via dimension-wise summation: each parameter tensor is reduced along all axes except the last, yielding a row vector that is then concatenated across all layers into a single vector per task.
This produces a compressed vector of dimension $D_r \approx 10^6$, achieving approximately $1000\times$ reduction.

The per-task gradients are collected during the existing policy-optimization backward passes, so no additional backward pass is required.
The per-step overhead consists of two parts: gradient reduction by element-wise accumulation at $O(K \cdot |\theta|)$ (no matrix multiplications), and $K(K{-}1)$ pairwise cosine similarities on $D_r$-dimensional vectors at $O(K^2 \cdot D_r)$.
The dominant cost remains the forward-backward pass at $O(|\theta| \cdot L \cdot B)$ FLOPs; in our setting ($L{=}6144$, $B{=}128$), the overhead ratio is below $0.001\%$.

The approach scales gracefully with the number of tasks $K$.
The gradient reduction costs $O(|\theta|)$ per task and thus grows only linearly with K.
The pairwise similarity cost $O(K^2 \cdot D_r)$ operates on the compressed vectors; even at $K{=}100$, this amounts to $\sim 10^{10}$ FLOPs, negligible compared to the $\sim 10^{16}$ FLOPs of a single forward-backward pass.

To verify that the compressed vectors capture task relationships, we compare cosine similarities computed from compressed and full gradients on Qwen2.5-Coder-7B.
Since computing full-gradient similarities is expensive, we perform this comparison over the first 80 training iterations, yielding $80\times 6=480$ paired measurements across the six task pairs.
The two measures agree in sign for 443 of the 480 comparisons (92.3\%), with a Spearman correlation of 0.91 and a mean absolute error of 0.041.
The compressed similarities therefore provide nearly the same task-relationship signal as full-gradient similarities at a small fraction of the cost.

\subsection{Utility Estimation and Normalization Details}
\label{sec:appendix_scheduler}

\paragraph{EMA Update Rules.}
For task-level utilities, we use exponential moving averages with $\alpha=0.9$:
\begin{equation}
\mathsf{EMA}_{\alpha}\!\left(\mathcal{U}_{\mathsf{pot}}^{(t-1)}(T_k)\right)
= \alpha \cdot \mathcal{U}_{\mathsf{pot}}^{(t-1)}(T_k)
+ (1-\alpha)\cdot
\mathsf{EMA}_{\alpha}\!\left(\mathcal{U}_{\mathsf{pot}}^{(t-2)}(T_k)\right),
\end{equation}
\begin{equation}
\mathsf{EMA}_{\alpha}\!\left(\mathcal{U}_{\mathsf{syn}}^{(t-1)}(T_k)\right)
= \alpha \cdot \mathcal{U}_{\mathsf{syn}}^{(t-1)}(T_k)
+ (1-\alpha)\cdot
\mathsf{EMA}_{\alpha}\!\left(\mathcal{U}_{\mathsf{syn}}^{(t-2)}(T_k)\right).
\end{equation}
Similarly, for prompt-level utilities:
\begin{equation}
\mathsf{EMA}_{\alpha}\!\left(\mathcal{U}_{\mathsf{pot}}^{(t-1)}(q,T_k)\right)
= \alpha \cdot \mathcal{U}_{\mathsf{pot}}^{(t-1)}(q,T_k)
+ (1-\alpha)\cdot
\mathsf{EMA}_{\alpha}\!\left(\mathcal{U}_{\mathsf{pot}}^{(t-2)}(q,T_k)\right),
\end{equation}
\begin{equation}
\mathsf{EMA}_{\alpha}\!\left(\mathcal{U}_{\mathsf{prog}}^{(t-1)}(q,T_k)\right)
= \alpha \cdot \mathcal{U}_{\mathsf{prog}}^{(t-1)}(q,T_k)
+ (1-\alpha)\cdot
\mathsf{EMA}_{\alpha}\!\left(\mathcal{U}_{\mathsf{prog}}^{(t-2)}(q,T_k)\right).
\end{equation}

\paragraph{Normalization Procedures.}
We normalize $\mathsf{EMA}_{\alpha}\!\left(\mathcal{U}_{\mathsf{pot}}^{(t-1)}(T_k)\right)$ to $[0,1]$ using min-max normalization:
\begin{equation}
\mathsf{EMA}_{\alpha}\!\left(\mathcal{U}_{\mathsf{pot}}^{(t-1)}(T_k)\right)
\leftarrow
\frac{
\mathsf{EMA}_{\alpha}\!\left(\mathcal{U}_{\mathsf{pot}}^{(t-1)}(T_k)\right)
- \min_j \mathsf{EMA}_{\alpha}\!\left(\mathcal{U}_{\mathsf{pot}}^{(t-1)}(T_j)\right)}
{\max_j \mathsf{EMA}_{\alpha}\!\left(\mathcal{U}_{\mathsf{pot}}^{(t-1)}(T_j)\right)
- \min_j \mathsf{EMA}_{\alpha}\!\left(\mathcal{U}_{\mathsf{pot}}^{(t-1)}(T_j)\right)},
\end{equation}
A small constant $\epsilon=10^{-8}$ is added to the denominator to prevent division by zero.
For $\mathsf{EMA}_{\alpha}\!\left(\mathcal{U}_{\mathsf{syn}}^{(t-1)}(T_k)\right)$, we normalize to $[-1,1]$:
\begin{equation}
\mathsf{EMA}_{\alpha}\!\left(\mathcal{U}_{\mathsf{syn}}^{(t-1)}(T_k)\right)
\leftarrow
2 \cdot
\frac{
\mathsf{EMA}_{\alpha}\!\left(\mathcal{U}_{\mathsf{syn}}^{(t-1)}(T_k)\right)
- \min_j \mathsf{EMA}_{\alpha}\!\left(\mathcal{U}_{\mathsf{syn}}^{(t-1)}(T_j)\right)}
{\max_j \mathsf{EMA}_{\alpha}\!\left(\mathcal{U}_{\mathsf{syn}}^{(t-1)}(T_j)\right)
- \min_j \mathsf{EMA}_{\alpha}\!\left(\mathcal{U}_{\mathsf{syn}}^{(t-1)}(T_j)\right)}
- 1.
\end{equation}
Both normalizations are performed across tasks at each step and are therefore relative: for synergy, the task with the lowest value is mapped to $-1$ even if all raw cosine similarities are positive, so a negative normalized value indicates relatively low synergy rather than an actual gradient conflict.

\paragraph{Gradient Estimation.}
For each task $T_k$, we compute $\mathbf{g}_k$ by averaging gradients over the current mini-batch.
To reduce noise in synergy estimation, we maintain gradient EMAs:
\begin{equation}
\bar{\mathbf{g}}_k(t) = \alpha \cdot \mathbf{g}_k(t) + (1-\alpha) \cdot \bar{\mathbf{g}}_k(t-1).
\end{equation}
Cross-task synergy is then computed from these smoothed gradients using the compression procedure described in Appendix~\ref{sec:appendix_gradient_compression}.

\paragraph{Quota Rounding.}
Since $N_k^{(t)}$ computed via softmax may be non-integer, we round quotas while preserving the total batch size $B$.
We use stochastic rounding: for each task, we set $N_k^{(t)} = \lfloor N_k^{(t)} \rfloor$ with probability $1 - (N_k^{(t)} - \lfloor N_k^{(t)} \rfloor)$ and $N_k^{(t)} = \lceil N_k^{(t)} \rceil$ otherwise.
We adjust the last task's quota to ensure $\sum_k N_k^{(t)} = B$ exactly.

\paragraph{Prompt Weight Initialization.}
Every prompt starts with a sampling weight of 1, the upper bound of $\sigma(\cdot)$.
Once a prompt has been selected and its utility computed, its weight is updated to $\sigma\big(\mathcal{U}^{(t)}(q,T_k)\big)<1$.
Prompts that have not yet been selected therefore keep the largest weight and are preferred for sampling.
This encourages coverage of each task's dataset but does not guarantee it.

\subsection{Reward Function Implementation Details}
\label{sec:appendix_reward}

\paragraph{Code I/O Prediction.}
Given a function signature and docstring, the model predicts:
(1) output for a given input, or (2) input that produces a specified output.

The reward is defined as:
\begin{equation}
R_{\text{IO}}(o) = 
\begin{cases}
-0.1, & \text{if } o \text{ has wrong format}, \\
Exec(o), & \text{otherwise},
\end{cases}
\end{equation}
where $Exec(o)\in\{0,1\}$ is an execution-based correctness metric.

\textit{Implementation:}
For output prediction, we construct: \texttt{assert f(input) == predicted\_output}.
For input prediction: \texttt{assert f(predicted\_input) == output}.
We execute the assertion in a sandboxed environment (Docker container with resource limits) with a 5-second timeout.
If execution succeeds without errors or timeouts, $Exec(o)=1$; otherwise $Exec(o)=0$.

\paragraph{Code Generation.}
Given a natural-language problem description, the model generates a complete function implementation.

The reward is:
\begin{equation}
R_{\text{CG}}(o) =
\begin{cases}
-0.1, & \text{if } o \text{ has wrong format}, \\
PassRatio(o), & \text{otherwise},
\end{cases}
\end{equation}
where $PassRatio(o) = \frac{\#\text{tests passed}}{\#\text{total tests}}$.

\textit{Implementation:}
We execute all provided unit tests against the generated code.
Each test is run in isolation with a 10-second timeout.
Tests that timeout or raise exceptions are counted as failures.
$PassRatio(o)$ is the fraction of tests that pass.

\paragraph{Unit Test Generation.}
Given a focal function and project context, the model generates test cases.

The reward is:
\begin{equation}
R_{\text{UT}}(o) =
\begin{cases}
-0.1, & \text{if } o \text{ has wrong format}, \\
Coverage(o), & \text{otherwise},
\end{cases}
\end{equation}
where $Coverage(o) = \frac{LineCov(o) + BranchCov(o)}{2}$.

\textit{Implementation:}
We use JaCoCo (Java Code Coverage Library) for coverage measurement.
Generated tests are first compiled with \texttt{javac}.
If compilation fails, $Coverage(o)=0$.
Otherwise, we execute tests with JaCoCo agent enabled, which instruments bytecode to track coverage.
$LineCov(o)$ is the percentage of executable lines covered, and $BranchCov(o)$ is the percentage of conditional branches covered.
Both metrics are normalized to $[0,1]$ and averaged.

\paragraph{Commit Message Generation.}
Given a code diff, the model generates a descriptive commit message.

The reward is:
\begin{equation}
R_{\text{CM}}(o) =
\begin{cases}
-0.1, & \text{if } o \text{ has wrong format}, \\
TextSim(o,gt), & \text{otherwise},
\end{cases}
\end{equation}
where $gt$ is the ground-truth message and $TextSim(o,gt) = \frac{BLEU(o,gt) + ROUGE(o,gt) + METEOR(o,gt)}{3}$.

\textit{Implementation:}
We compute BLEU-4 with smoothing using SacreBLEU.
ROUGE-L is computed using the longest common subsequence F1 score from the \texttt{rouge-score} library.
METEOR is computed using the NLTK implementation with English stemming and WordNet synonyms.
All scores are normalized to $[0,1]$ and averaged.

\paragraph{Format Checking.}
All tasks use a structured prompt template requiring output format:
\texttt{<think>...</think> <answer>...</answer>}.

We use regular expressions to check format compliance:
\begin{verbatim}
pattern = r'<think>(.+?)</think>\s*<answer>(.+?)</answer>'
\end{verbatim}
If the pattern matches and both tags contain non-empty content, $R_{\text{format}}(o)=1$; otherwise $R_{\text{format}}(o)=0$.

The final reward for task $k$ combines task-specific and format rewards:
\begin{equation}
R_k(o) = 0.5 \cdot R_{\text{task},k}(o) + 0.5 \cdot R_{\text{format}}(o),
\end{equation}
where $R_{\text{task},k} \in \{R_{\text{IO}}, R_{\text{CG}}, R_{\text{UT}}, R_{\text{CM}}\}$.
The $-0.1$ penalty for malformed outputs follows prior RL work on LLM agents, and the equal weighting of task and format rewards is our own design choice. Both settings are specific to our current experiments and can be adjusted for new tasks without changing {\tool}.



\subsection{Prompt Templates}
\label{sec:appendix_prompts}

All tasks use structured prompts that guide the model to first reason in \texttt{<think>} tags, then provide the answer in \texttt{<answer>} tags.
This format encourages chain-of-thought reasoning and facilitates reward computation.

\begin{promptbox}{Code Input Prediction Prompt Template}
\textless\textbar im\_start\textbar\textgreater \space system\\
You are a helpful assistant. The assistant first thinks
about the reasoning process in the mind and then provides
the user with the answer. The reasoning process and answer
are enclosed within \textless think\textgreater \space \textless/think\textgreater \space and \textless answer\textgreater \space \textless/answer\textgreater \space
tags, respectively.\\
\textless\textbar im\_end \textbar \textgreater\\
\textless\textbar im\_start\textbar\textgreater user\\
You are given a function f and an output in the form
f(??) == output. Your task is to find any input such that
executing f on the input leads to the given output. There
may be multiple answers, but only output one.
After thinking, when you finally reach an answer, clearly
provide your answer as a passing assertion containing the
input and the given output within \textless answer\textgreater \space \textless/answer\textgreater \space tags.
e.g., \textless answer\textgreater \space assert f(16) == 17 \textless/answer\textgreater.\\

Given function:\\
\redvar{python\_code}\\
assert f(??) == \redvar{output}\\
\textless\textbar im\_end \textbar\textgreater\\
\textless\textbar im\_start\textbar\textgreater assistant.
\end{promptbox}

\begin{promptbox}{Code Output Prediction Prompt Template}
\textless\textbar im\_start\textbar\textgreater \space system\\
You are a helpful assistant. The assistant first thinks
about the reasoning process in the mind and then provides
the user with the answer. The reasoning process and answer
are enclosed within \textless think\textgreater \space \textless/think\textgreater \space and \textless answer\textgreater \space \textless/answer\textgreater \space
tags, respectively.\\
\textless\textbar im\_end\textbar\textgreater\\
\textless\textbar im\_start\textbar\textgreater \space user\\
You are given a Python function and an assertion containing
an input to the function. Complete the assertion with a
literal (no unsimplified expressions, no function calls)
containing the output based on the logical reasoning of
given function.\\
After thinking, when you finally reach an answer, clearly
provide your answer as the full assertion with the correct
output within \textless answer\textgreater \space \textless/answer\textgreater \space tags.\\
e.g., \textless answer\textgreater \space assert f("hi") == "bhihia" \textless/answer\textgreater.\\

Given function:\\
\redvar{python\_code}\\
assert f(\redvar{input}) == ??\\
\textless\textbar im\_end\textbar\textgreater\\
\textless\textbar im\_start\textbar\textgreater \space assistant
\end{promptbox}

\begin{promptbox}{Code Generation Prompt Template}
\textless\textbar im\_start\textbar\textgreater \space system\\
You are a helpful assistant. The assistant first thinks
about the reasoning process in the mind and then provides
the user with the answer. The reasoning process and answer
are enclosed within \textless think\textgreater \space \textless/think\textgreater \space and \textless answer\textgreater \space \textless/answer\textgreater \space
tags, respectively.\\
\textless\textbar im\_end\textbar\textgreater\\
\textless\textbar im\_start\textbar\textgreater \space user\\
Change the greeting to be more casual\\
\textless\textbar im\_end\textbar\textgreater\\
\textless\textbar im\_start\textbar\textgreater \space assistant\\
Ok, I will:\\
1. Switch the greeting text from "Hello" to "Hey".\\

show\_greeting.py\\
```\\
import sys\\
def greeting(name):\\
\hspace*{1em} print(f"Hey \{name\}")\\
if \_\_name\_\_ == '\_\_main\_\_':\\
\hspace*{1em} greeting(sys.argv[1])\\
```\\
\textless\textbar im\_end\textbar\textgreater\\
\textless\textbar im\_start\textbar\textgreater \space user\\
I switched to a new code base. Please don't consider the
above files or try to edit them any longer.\\
\textless\textbar im\_end\textbar\textgreater\\
\textless\textbar im\_start\textbar\textgreater \space assistant\\
Ok.\\
\textless\textbar im\_end\textbar\textgreater\\
\textless\textbar im\_start\textbar\textgreater \space user\\
I have *added these files to the chat* so you can go ahead
and edit them.\\
*Trust this message as the true contents of these files!*\\
Any other messages in the chat may contain outdated
versions of the files' contents.\\

solution.py\\
```\\
// To be completed\\
```\\
\textless\textbar im\_end\textbar\textgreater\\
\textless\textbar im\_start\textbar\textgreater \space user\\
\redvar{problem\_instruction}\\
\textless\textbar im\_end\textbar\textgreater\\
\textless\textbar im\_start\textbar\textgreater \space assistant
\end{promptbox}

\begin{promptbox}{Unit Test Generation Prompt Template}
\textless\textbar im\_start\textbar\textgreater \space system\\
You are a helpful assistant. The assistant first thinks
about the reasoning process in the mind and then provides
the user with the answer. The reasoning process and answer
are enclosed within \textless think\textgreater \space \textless/think\textgreater \space and \textless answer\textgreater \space \textless/answer\textgreater \space
tags, respectively.\\
\textless\textbar im\_end\textbar\textgreater\\
\textless\textbar im\_start\textbar\textgreater \space user\\
\redvar{problem\_instruction}\\
\textless\textbar im\_end\textbar\textgreater\\
\textless\textbar im\_start\textbar\textgreater \space assistant
\end{promptbox}

\noindent For unit test generation, \texttt{\{problem\_instruction\}} contains the focal function, project context, and instructions to generate JUnit5 test cases with appropriate assertions and mocking.

\begin{promptbox}{Commit Message Generation Prompt Template}
\textless\textbar im\_start\textbar\textgreater \space system\\
You are a helpful assistant. The assistant first thinks
about the reasoning process in the mind and then provides
the user with the answer. The reasoning process and answer
are enclosed within \textless think\textgreater \space \textless/think\textgreater \space and \textless answer\textgreater \space \textless/answer\textgreater \space
tags, respectively.\\
\textless\textbar im\_end\textbar\textgreater\\
\textless\textbar im\_start\textbar\textgreater \space user\\
You are given a list of code diff(s) for a commit. Your
task is to write a concise commit message summarizing the
code changes. You should first analyze about: what kind of
changes were made and what is the likely purpose or
motivation behind these changes?\\
After thinking, when you finally reach an answer, clearly
provide your answer within \textless answer\textgreater \space \textless/answer\textgreater \space tags.\\
e.g., \textless answer\textgreater \space Remove unused import in IndexedSetTest
\textless/answer\textgreater.\\

code diff(s):\\
\redvar{diffcode}\\
\textless\textbar im\_end\textbar\textgreater\\
\textless\textbar im\_start\textbar\textgreater \space assistant
\end{promptbox}

\subsection{Detailed Algorithm Pseudocode}
\label{sec:appendix_detailed_algorithm}

Algorithm~\ref{alg:detailed_mtrl} provides a more detailed version of the training procedure, including low-level implementation details omitted from the main text.

\begin{algorithm}[]
\caption{Detailed Training Procedure for \tool}
\label{alg:detailed_mtrl}
\begin{algorithmic}[1]
\REQUIRE Tasks $\T=\{T_1,\dots,T_K\}$, batch size $B$, group size $G$, total iterations $M$
\REQUIRE EMA coefficient $\alpha$, temperature $\tau$, adjustment rate $\lambda$
\REQUIRE Base KL coefficients $\{\beta_{\text{base},k}\}_{k=1}^K$
\STATE Initialize policy $\pi_\theta$, reference policy $\pi_{\text{ref}}$;
\STATE Initialize EMA states for task utilities $U_k^{\mathrm{pot}}$, $U_k^{\mathrm{syn}}$ and prompt utilities $U_{q,k}^{\text{pot}}$, $U_{q,k}^{\text{prog}}$;
\STATE Initialize gradient EMAs: $\bar{\grad}_k(0) \leftarrow 0$ for all $k$
\FOR{$t=1$ to $M$}
    \STATE \textit{\textcolor{teal}{// Phase 1: Utility Estimation}}
    \FOR{each task $T_k$}
        \STATE Compute instantaneous utilities $U_k^{\mathrm{pot}}(t-1)$, $U_k^{\mathrm{syn}}(t-1)$ from iteration $t-1$;
        \STATE Update $\mathsf{EMA}_{\alpha}(U_k^{\mathrm{pot}}(t-1))$ and $\mathsf{EMA}_{\alpha}(U_k^{\mathrm{syn}}(t-1))$;
    \ENDFOR
    \STATE Normalize $\{\mathsf{EMA}_{\alpha}(U_k^{\mathrm{pot}}(t-1))\}$ to $[0,1]$ and $\{\mathsf{EMA}_{\alpha}(U_k^{\mathrm{syn}}(t-1))\}$ to $[-1,1]$
    \STATE \textit{\textcolor{teal}{// Phase 2: Task Quota Allocation}}
    \FOR{each task $T_k$}
        \STATE Compute combined utility: $U_k(t) \leftarrow \mathsf{EMA}_{\alpha}(U_k^{\mathrm{pot}}(t-1)) + \mathsf{EMA}_{\alpha}(U_k^{\mathrm{syn}}(t-1))$;
    \ENDFOR
    \STATE Allocate quotas: $N_k^{(t)} \leftarrow B \cdot \frac{\exp(U_k(t)/\tau)}{\sum_j \exp(U_j(t)/\tau)}$ for all $k$;
    \STATE Round quotas to integers while preserving $\sum_k N_k^{(t)} = B$;
    \STATE \textit{\textcolor{teal}{// Phase 3: Prompt Selection}}
    \FOR{each task $T_k$}
        \FOR{each prompt $q \in \mathcal{D}_k$}
            \STATE Update $\mathsf{EMA}_{\alpha}(U_{q,k}^{\text{pot}}(t-1))$ and $\mathsf{EMA}_{\alpha}(U_{q,k}^{\text{prog}}(t-1))$;
            \STATE Compute $U_{q,k}(t) \leftarrow \mathsf{EMA}_{\alpha}(U_{q,k}^{\text{pot}}(t-1)) + \mathsf{EMA}_{\alpha}(U_{q,k}^{\text{prog}}(t-1))$;
        \ENDFOR
        \STATE Sample $N_k^{(t)}$ prompts from $\mathcal{D}_k$ using weights $w_{q,k}(t) = \mathrm{sigmoid}(U_{q,k}(t))$ ($w_{q,k}(t)=1$ if $q$ was never selected) to form prompt batch $\mathcal{Q}_k^{(t)}$;
    \ENDFOR
    \STATE Form mixed batch: $\mathcal{Q}^{(t)} \leftarrow \bigcup_{k=1}^K \mathcal{Q}_k^{(t)}$;
    \STATE \textit{\textcolor{teal}{// Phase 4: Rollout Generation and Reward Computation}}
    \STATE Save old policy: $\pi_{\theta_{\text{old}}} \leftarrow \pi_\theta$;
    \FOR{each prompt $q \in \mathcal{Q}^{(t)}$ from task $T_k$}
        \STATE Sample $G$ rollouts: $\{o_i\}_{i=1}^G \sim \pi_{\theta_{\text{old}}}(\cdot \mid q)$;
        \STATE Compute rewards: $\{r_i = R_k(q,o_i)\}_{i=1}^G$;
        \STATE Store $(q, \{o_i\}, \{r_i\}, k)$ in batch buffer;
    \ENDFOR
    \STATE \textit{\textcolor{teal}{// Phase 5: Policy Optimization}}
    \FOR{each task $T_k$}
        \STATE Set KL coefficient: $\beta_k(t) \leftarrow \beta_{\text{base},k} \cdot (1 + \lambda \cdot U_k(t))$
    \ENDFOR
    \STATE Compute per-task GRPO objectives $\{J_k(\theta;t)\}$ with $\{\beta_k(t)\}$;
    \STATE Compute multi-task objective: $J_{\text{MT}}(\theta;t) \leftarrow \sum_k \frac{N_k^{(t)}}{B} J_k(\theta;t)$;
    \STATE Update policy: $\theta \leftarrow \theta + \eta \nabla_\theta J_{\text{MT}}(\theta;t)$ (via AdamW);
    \STATE \textit{\textcolor{teal}{// Phase 6: Utility Update for Next Iteration}}
    \FOR{each task $T_k$}
        \STATE Take the gradient $\grad_k(t)$ of $J_k(\theta;t)$ from the Phase~5 backward pass;
        \STATE Update gradient EMA: $\bar{\grad}_k(t) \leftarrow \alpha \grad_k(t) + (1-\alpha)\bar{\grad}_k(t-1)$;
        \STATE Compute $U_k^{\mathrm{pot}}(t)$ from reward variances;
        \STATE Compute $U_k^{\mathrm{syn}}(t)$ from $\{\bar{\grad}_j(t)\}_{j=1}^K$;
    \ENDFOR
    \FOR{each prompt $q$ sampled at iteration $t$}
        \STATE Update $U_{q,k}^{\text{pot}}(t)$ and $U_{q,k}^{\text{prog}}(t)$ from rollout rewards;
    \ENDFOR
\ENDFOR

\textbf{Return} $\pi_\theta$;
\end{algorithmic}
\end{algorithm}

\section{Experimental Setup Details}
\label{sec:appendix_implementation}

\subsection{Dataset Statistics}
\label{sec:appendix_dataset}

We describe the construction and format of each task dataset below.

\paragraph{Code I/O Prediction.} We collect 2,088 training samples from LeetCode~\cite{exp-data-LeetCodeDataset}. Each sample provides a Python function signature along with a docstring, and the model is asked to either predict the output given a specific input, or find a valid input that produces a given output.

\paragraph{Code Generation.} We collect 2,082 training samples from LiveCodeBench~\cite{exp-data-lcb} and Codeforces~\cite{exp-data-Codeforces}. Each sample contains a natural-language problem description, and the model is required to generate a complete, executable solution.

\paragraph{Unit Test Generation.} We extract 1,148 Java functions from real-world open-source repositories covering mainstream frameworks (e.g., MyBatis, Spring, Flux), and standardize their test environments to JUnit5 + Mockito to enable executable validation during training. Each sample provides the target function along with its project context, and the model is asked to generate a set of unit tests that achieves high line and branch coverage.

\paragraph{Commit Message Generation.} We collect 2,446 training samples from the MCMD dataset~\cite{exp-data-MCMD}, filtering out messages shorter than 5 words, non-Java code changes, and uninformative entries. Each sample provides a code diff from a real-world commit, and the model is asked to generate a concise and descriptive commit message.

\subsection{Evaluation Benchmarks and Metrics}
\label{sec:appendix_benchmarks}

\paragraph{CRUXEval.}
CRUXEval~\cite{exp-benchmark-crux} is a benchmark of 800 short Python functions (3--13 lines) designed to evaluate code reasoning and execution understanding.
It consists of two complementary subtasks: \textit{input prediction} (CRUXEval-I), where the model is given a function and an expected output and must find a valid input; and \textit{output prediction} (CRUXEval-O), where the model is given a function and an input and must predict the resulting output.
Both subtasks are evaluated by executing the model's answer and checking whether the assertion holds, yielding a binary correctness signal.
We report \textit{input accuracy} and \textit{output accuracy} as the pass@1 rates on CRUXEval-I and CRUXEval-O, respectively.

\paragraph{Aider-Polyglot-Python.}
Aider-Polyglot-Python~\cite{exp-benchmark-Aider} is a subset of the Aider polyglot benchmark, consisting of 135 challenging coding exercises drawn from Exercism.
Each exercise provides a natural-language specification and a set of hidden unit tests; the model must generate a complete Python solution that passes all tests.
The benchmark is specifically designed to remain challenging for frontier models, as it selects only the hardest exercises unsolved by most tested models.
We report \textit{pass@1} (the fraction of problems solved on the first attempt) and \textit{pass@2} (the fraction solved within two attempts).

\paragraph{Defects4J 2.0.}
Defects4J 2.0~\cite{exp-benchmark-defect4j} is a curated collection of 737 reproducible bugs from 17 real-world open-source Java projects, widely used for evaluating automated test generation.
Given a target Java method and its project context, the model generates a test suite that is then compiled and executed against the project.
We report three metrics: \textit{line coverage} (percentage of executable lines exercised by the generated tests), \textit{branch coverage} (percentage of conditional branches covered), and \textit{compilation rate} (percentage of generated test suites that successfully compile), measured using the JaCoCo coverage tool.

\paragraph{MCMDEval+.}
MCMDEval+~\cite{exp-data-MCMD} is an evaluation set derived from the MCMD dataset with 500 samples.
Given a code diff, the model generates a commit message that is compared against the human-written reference.
We report three standard text generation metrics: \textit{BLEU}~\cite{exp-metric-BLEU} (n-gram precision with a brevity penalty), \textit{ROUGE-L}~\cite{exp-metric-rouge} (F1 score based on the longest common subsequence), and \textit{METEOR}~\cite{exp-metric-ETEOR} (harmonic mean of precision and recall with stemming and synonym matching).

\subsection{Baseline Configurations}
\label{sec:appendix_baselines}

\paragraph{Single-Task Learning (STL).}
We train a separate GRPO~\cite{grpo} model for each of the four tasks independently. Each specialist is trained with the same hyperparameters as {\tool}, using a task-specific KL coefficient chosen via validation. STL serves as the per-task specialist reference: each model excels on its own task but falls short of the other specialists on their tasks.

\paragraph{Joint Learning.}
Following the experimental setup of Wu et al.~\cite{intro-MTRL-2}, we mix all four tasks uniformly within each training batch (equal quota per task) and train a single shared model using GRPO.
Wu et al. also compare uniform sampling with sampling tasks in proportion to their gradient magnitudes and find no consistent advantage for the latter, since larger gradients do not reliably indicate larger learning gains. We therefore use uniform mixing as the representative joint-learning baseline.
The KL coefficient $\beta{=}5\text{e-}3$ is selected as the best uniform value via preliminary experiments across $\{10^{-4}, 5\times10^{-3}, 10^{-2}\}$.

\paragraph{Curriculum Learning.}
Following the curriculum-based MTRL paradigm of OMNI-THINKER~\cite{intro-MTRL-3}, we train tasks sequentially in a fixed order: Code I/O $\to$ Code Generation $\to$ Unit Test $\to$ Commit Message. This order follows the software development workflow in Figure~\ref{fig:development} (understanding, implementation, testing, documentation) rather than task difficulty. The model is carried forward across stages without resetting.

\paragraph{Model Merging.}
We first train four single-task GRPO specialists (identical to STL above) and select the best checkpoint per task.
We then merge the four checkpoints using TIES-Merging~\cite{TIES} with scaling factor $\lambda{=}1.0$, which resolves parameter sign conflicts and prunes redundant updates before averaging.
This baseline represents post-hoc integration without any joint optimization.

\subsection{Hyperparameter Configurations}
\label{sec:appendix_hyperparameters}

\paragraph{Training and RL Settings.}
All methods share the same general training configuration: learning rate $5\text{e-}7$ with cosine scheduler, AdamW optimizer, gradient clip $1.0$, global batch size $128$, $8$ rollouts per sample, rollout temperature $1.0$, top-p $1.0$, top-k $50$, max prompt length $4{,}096$, max response length $2{,}048$, trained for $400$ steps on $32$ Ascend 910B-B3 NPUs.

\paragraph{{\tool}-specific Settings.}
We use EMA coefficient $\alpha{=}0.9$ for utility estimation at both task and prompt levels to smooth noisy per-batch statistics.
For task allocation, we set softmax temperature $\tau{=}1.0$: lower values (e.g., $\tau{=}0.5$) over-concentrate on the highest-utility tasks and starve others; higher values (e.g., $\tau{=}2.0$) approach uniform sampling and fail to exploit utility differences.
For dynamic KL regularization, we set adjustment rate $\lambda{=}0.2$, which is robust across a moderate range ($\lambda \in [0.1, 0.4]$); too small approximates static KL, while too large suppresses beneficial policy updates.

\paragraph{Task-Specific KL Base Coefficients.}
We assign $\beta{=}10^{-2}$ to tasks that must output complete executable programs (code generation, unit test) to preserve syntactic correctness, and $\beta{=}10^{-4}$ to tasks whose outputs are short answers or natural language (code I/O, commit message) to allow expressive reasoning.
For baselines requiring a uniform $\beta$, we select the best value via preliminary experiments on Qwen2.5-Coder-7B (Table~\ref{tab:kl_selection}).

\begin{table*}[h]
\centering
\footnotesize
\caption{Effect of uniform KL coefficient $\beta$ on Joint Learning (Qwen2.5-Coder-7B). No single value resolves the conflict between task types; $\beta{=}5\text{e-}3$ is the best compromise and is used for all uniform-$\beta$ baselines.}
\label{tab:kl_selection}
\begin{tabular}{lccccc}
\toprule
$\beta$ & Code I/O (In/Out) & Code Gen (P@1/P@2) & Unit Test (Ln/Br/Cmp) & Commit Msg (B/R/M) & Avg \\
\midrule
1e-4        & 51.5/53.4 & 25.9/37.0 & 19.4/16.2/41.5 & 33.8/20.4/18.6 & 31.77 \\
\textbf{5e-3} & \textbf{51.4/54.0} & \textbf{32.6/43.0} & \textbf{25.9/21.0/51.1} & 30.7/17.1/16.0 & \textbf{34.28} \\
1e-2        & 47.5/49.0 & 37.0/43.7 & 25.2/23.2/53.2 & 25.6/13.4/12.2 & 33.00 \\
\bottomrule
\end{tabular}
\end{table*}

$\beta{=}10^{-4}$ favors commit message generation but severely hurts code generation and unit test generation; $\beta{=}10^{-2}$ does the opposite.
Since all baselines use their respective best-performing KL coefficients, the consistent gains of {\tool} confirm that improvements stem from the scheduling and dynamic regularization modules, not from the base KL configuration.

\section{Additional Experimental Results}
\label{sec:appendix_additional}

\subsection{Semantic Evaluation of Commit Messages}
\label{sec:appendix_llm_judge}
To complement the lexical-overlap metrics, we evaluate the generated commit messages on MCMDEval+ with an LLM judge following Zeng et al.~\cite{cmg-llm-judge}.
We use GPT-4 under the chain-of-thought (CoT) and few-shot setting, which Zeng et al. report as achieving the strongest agreement with human judgment.
The judge receives the code diff and the generated message and assigns a score from 0 to 4 based on the two criteria of Zeng et al.: whether the message correctly describes \textit{what} changed, and whether it correctly conveys \textit{why} the change was made.
A score of 4 requires both to be accurate and complete, and a score of 0 indicates that the message is unrelated to the diff.
Table~\ref{tab:llm_judge} reports the average score.

{\tool} achieves the highest score on both base models, ahead of the commit-message specialist and all multi-task baselines, so its improvement holds beyond lexical overlap.
On Qwen2.5-Coder-7B, STL (Commit Msg) attains a higher BLEU than {\tool} (41.2 vs.\ 35.2; Table~\ref{tab:main_results}) yet a lower judge score (2.88 vs.\ 3.08), indicating that a higher overlap with the reference does not necessarily yield a more accurate description of the change. Table~\ref{tab:cmg_comparison} in Appendix~\ref{sec:appendix_examples} shows a representative example.

\begin{table}[ht]
\centering
\small
\caption{LLM-as-a-judge scores (0--4) for commit message generation on MCMDEval+. Bolded values mark the best per column; underscored values mark the second best.}
\label{tab:llm_judge}
\begin{tabular}{lcc}
\toprule
Method & Qwen2.5-Coder-7B & Qwen3-8B \\
\midrule
Base Model & 1.62 & 2.05 \\
STL (Code I/O) & 2.06 & 2.25 \\
STL (Code Gen) & 2.10 & 2.29 \\
STL (Unit Test) & 1.86 & 2.10 \\
STL (Commit Msg) & \underline{2.88} & \underline{2.93} \\
\midrule
Joint Learning & 2.18 & 2.50 \\
Curriculum Learning & 2.22 & 2.42 \\
Model Merging & 2.26 & 2.47 \\
\midrule
{\tool} & \textbf{3.08} & \textbf{3.20} \\
\bottomrule
\end{tabular}
\end{table}

\section{Qualitative Examples}

We compare the outputs of Joint Learning and ASTOR on representative examples of code output prediction, code generation, and commit message generation, respectively.

\label{sec:appendix_examples}











\begin{table}[!htbp]
\centering\scriptsize
\caption{Comparison of Output Prediction. Baseline answer in \textcolor{deep_red}{\textbf{red}}; {\tool} answer in \textcolor{deep_green}{\textbf{green}}.}
\label{tab:output_prediction_comparison_new}

\begin{tcolorbox}[colback=gray!6,colframe=gray!50,title=\textbf{Prompt},fonttitle=\bfseries\small,left=4pt,right=4pt,top=2pt,bottom=2pt]
Given a Python function and an assertion, complete the assertion with the output.

\smallskip\textbf{Function:}

\texttt{def f(text):}\\
\mytab\texttt{text = text.upper()}\\
\mytab\texttt{count\_upper = 0}\\
\mytab\texttt{for char in text:}\\
\mytab\mytab\texttt{if char.isupper():}\\
\mytab\mytab\mytab\texttt{count\_upper += 1}\\
\mytab\mytab\texttt{else:}\\
\mytab\mytab\mytab\texttt{return 'no'}\\
\mytab\texttt{return count\_upper // 2}

\smallskip\blueback{assert f('ax') == ??}
\end{tcolorbox}

\medskip\textbf{Model Responses}

\smallskip\textbf{Baseline (Joint Learning):}
\begin{badbox}
<think>
The function returns 'no' if any character is not uppercase. Input 'ax' has lowercase letters.
</think>

\textbf{\textcolor{deep_red}{<answer> assert f('ax') == 'no' </answer>}}
\end{badbox}

\vspace{0.6em}\textbf{{\tool}:}
\begin{goodbox}
<think>
'ax'.upper() = 'AX'. Both chars are uppercase, count\_upper = 2, return 2 // 2 = 1.
</think>

\textbf{\textcolor{deep_green}{<answer> assert f('ax') == 1 </answer>}}
\end{goodbox}
\end{table}

\begin{table}[!htbp]
\centering\scriptsize
\caption{Comparison of Code Generation. Baseline answer in \textcolor{deep_red}{\textbf{red}}; {\tool} answer in \textcolor{deep_green}{\textbf{green}}.}
\label{tab:code_gen_comparison}

\begin{tcolorbox}[colback=gray!6,colframe=gray!50,title=\textbf{Prompt},fonttitle=\bfseries\small,left=4pt,right=4pt,top=2pt,bottom=2pt]
Implement a binary search algorithm in \texttt{binary\_search.py}. Use the \texttt{raise} statement to throw a \texttt{ValueError} when the value is not found.
\end{tcolorbox}

\medskip\textbf{Model Responses}

\smallskip\textbf{Baseline (Joint Learning):}
\begin{badbox}
<think>
Implement binary search returning -1 if not found.
</think>

\textbf{\textcolor{deep_red}{<answer>}}\\
\texttt{def find(search\_list, value):}\\
\mytab\texttt{low = 0}\\
\mytab\texttt{high = len(search\_list) - 1}\\
\mytab\texttt{while low <= high:}\\
\mytab\mytab\texttt{mid = (low + high) // 2}\\
\mytab\mytab\texttt{if search\_list[mid] == value: return mid}\\
\mytab\mytab\texttt{elif search\_list[mid] > value: high = mid - 1}\\
\mytab\mytab\texttt{else: low = mid + 1}\\
\mytab\texttt{return -1}\quad\textbf{\textcolor{deep_red}{\# wrong: should raise ValueError}}
\textbf{\textcolor{deep_red}{</answer>}}
\end{badbox}

\vspace{0.6em}\textbf{{\tool}:}
\begin{goodbox}
<think>
Implement binary search raising ValueError when not found.
</think>

\textbf{\textcolor{deep_green}{<answer>}}\\
\texttt{def find(search\_list, value):}\\
\mytab\texttt{low = 0}\\
\mytab\texttt{high = len(search\_list) - 1}\\
\mytab\texttt{while low <= high:}\\
\mytab\mytab\texttt{mid = (low + high) // 2}\\
\mytab\mytab\texttt{if search\_list[mid] == value: return mid}\\
\mytab\mytab\texttt{elif search\_list[mid] > value: high = mid - 1}\\
\mytab\mytab\texttt{else: low = mid + 1}\\
\mytab\texttt{raise ValueError("value not in array")}
\textbf{\textcolor{deep_green}{</answer>}}
\end{goodbox}
\end{table}

\begin{table}[!htbp]
\centering\scriptsize
\caption{Comparison of Commit Message Generation. Baseline answer in \textcolor{deep_red}{\textbf{red}}; {\tool} answer in \textcolor{deep_green}{\textbf{green}}.}
\label{tab:cmg_comparison}

\begin{tcolorbox}[colback=gray!6,colframe=gray!50,title=\textbf{Prompt},fonttitle=\bfseries\small,left=4pt,right=4pt,top=2pt,bottom=2pt]
Given the code diff below, write a concise commit message.

\smallskip\texttt{diff --git a/.../CoinmateAdapters.java}\\
\texttt{@@ -186,6 +186,7 @@ public class CoinmateAdapters \{}\\
\mytab\texttt{case "COMPLETED": status = COMPLETE; break;}\\
\texttt{+\mytab case "NEW":}\\
\mytab\texttt{case "SENT":}\\
\mytab\texttt{case "CREATED":}\\
\mytab\texttt{case "WAITING":}
\end{tcolorbox}

\medskip\textbf{Model Responses}

\smallskip\textbf{Baseline (Joint Learning):}
\begin{badbox}
<think>A new case for "NEW" status is added.</think>

\textbf{\textcolor{deep_red}{<answer> Add NEW status to CoinmateAdapters </answer>}}
\end{badbox}

\vspace{0.6em}\textbf{{\tool}:}
\begin{goodbox}
<think>The commit adds support for a new funding record status "NEW" in CoinmateAdapters.</think>

\textbf{\textcolor{deep_green}{<answer> Add support for NEW funding record status </answer>}}
\end{goodbox}
\end{table}

\section{Limitations and Future Work}
\label{sec:limit_future}

\paragraph{Task and Language Coverage}
The current experiments cover four coding tasks using Python and Java. While these tasks represent a reasonable cross-section of the development workflow, we have not yet validated {\tool} on other software engineering tasks such as code review or bug localization, nor on a wider set of programming languages. Extending {\tool} to more tasks and languages is a straightforward next step. Moreover, our setting focuses on single-turn tasks with verifiable rewards. Agentic scenarios such as GUI agent tasks involve multi-turn interaction and tool use, which our current setup does not cover. Domains outside coding, such as mathematical reasoning or general reinforcement learning from human feedback (RLHF), are likewise beyond the scope of this work. Extending utility-driven coordination to these settings is an important direction for future work.

\paragraph{Reward Quality for Generative Tasks}
For commit message generation, the reward is computed as an average of BLEU, ROUGE-L, and METEOR against reference messages. These metrics are convenient and verifiable, but they measure surface-level lexical overlap rather than semantic correctness or usefulness. Using a stronger reward signal for such tasks, for example, an LLM-based evaluator, could improve training quality and is worth exploring in future work.

\paragraph{Manual Setting of KL Base Coefficients}
The task-specific KL base coefficients $\beta_{\text{base},k}$ are set manually based on task type, with a higher value for code-generation tasks and a lower value for text-generation tasks. This works well in the current setting, but requires prior knowledge about each task's optimization characteristics. When applying {\tool} to new tasks, these coefficients would need to be re-tuned, and automating this process is left for future work.

\section{Broader Impacts}
\label{sec:broader_impacts}

\paragraph{Positive Impacts.}
{\tool} enables a single LLM to replace multiple task-specific RL specialists across the software development workflow, directly reducing memory and compute costs for AI-assisted coding deployments.
By improving model efficiency and multi-task capability, it lowers the barrier for practitioners and organizations to adopt capable coding assistants, potentially accelerating software development and improving code quality at scale.

\paragraph{Negative Impacts.}
Stronger automated code generation capabilities could be misused to produce malicious code, exploit software vulnerabilities, or automate cyberattacks.
We acknowledge this dual-use risk and note that responsible deployment of {\tool}-trained models should be accompanied by appropriate access controls and output filtering, consistent with the policies of the underlying base models (Qwen2.5-Coder and Qwen3).

\section{Extended Related Work}
\label{sec:extended_related_work}

\paragraph{Multi-Task Learning Foundations.}
Multi-task learning (MTL)~\cite{intro-MTRL-def} is the paradigm of training a single model on multiple related tasks simultaneously, leveraging shared structure to act as an inductive bias that improves generalization beyond what single-task models can achieve.
In deep learning, MTL is typically implemented via hard parameter sharing, where all tasks share hidden layers with task-specific output heads, or soft parameter sharing, where tasks maintain private parameters regularized toward a common solution.
In multi-task RL, representative architectures extend this with task-specific modules and routing networks, such as Soft Modularization~\cite{intro-MTRL-Tra-2}, which learns to compose shared modules dynamically, and approaches sharing experience via auxiliary tasks~\cite{intro-MTRL-Tra-1} to enrich representation learning.
A recurring finding across all these settings is that na\"{i}ve joint optimization creates dominance effects: tasks with larger gradients or shorter time horizons tend to monopolize training, degrading performance on others.
{\tool} operates in this same multi-task RL regime but specifically targets LLM post-training, where non-stationary online rollouts, heterogeneous reward semantics, and per-token KL regularization introduce qualitatively new challenges.

\paragraph{Gradient Manipulation in Multi-Task Learning.}
A prominent class of methods addresses task interference by modifying gradient directions or magnitudes at each update step.
PCGrad~\cite{related-PCGrad} detects conflicting task gradients (negative cosine similarity) and projects each gradient onto the normal plane of conflicting gradients, preventing destructive interference.
CAGrad~\cite{CAGrad} takes a principled optimization-theoretic view: it minimizes average loss while leveraging the worst local per-task improvement as a regularizer, provably converging to an average-loss minimum while outperforming PCGrad and MGDA on multi-task supervised and RL benchmarks.
GradDrop~\cite{GradDrop} probabilistically masks gradients at activation layers based on a sign-purity score, allowing gradients to compete while provably converging near joint local minima of all task losses.
While these methods are effective at gradient-level alignment, they operate entirely within a single update step and do not control which data is presented during that step, nor do they adapt the trust-region constraint (e.g., KL coefficient) to each task's training state.
{\tool}'s utility signal uses gradient cosine similarity as one component of cross-task synergy, but combines it with learning-potential signals to drive both data scheduling and adaptive KL regularization---decisions that gradient manipulation methods leave unaddressed.

\paragraph{Multi-Task RL for LLMs.}
With the rise of RL-based LLM post-training, several recent works have tackled multi-task settings.
Wu et al.~\cite{intro-MTRL-2} provide a systematic empirical analysis showing that na\"{i}ve gradient mixing in multi-task RL causes severe imbalance: tasks with inherently larger gradient norms dominate optimization regardless of their actual learning gains, a phenomenon not captured by standard training statistics such as reward or advantage.
Their findings motivate principled gradient-level corrections for LLMs post-training.
OMNI-THINKER~\cite{intro-MTRL-3} addresses cross-task forgetting using backward-transfer-guided task scheduling, where task sampling probabilities are adjusted according to backward transfer scores measured across previously encountered tasks; it also combines rule-based and preference-based reward signals in a hybrid supervision scheme.
While OMNI-THINKER introduces curriculum scheduling for LLM MTRL, its scheduling signal is based on backward transfer from a fixed task ordering rather than on dynamic real-time gradient interactions between tasks.
{\tool} estimates cross-task synergy online at each training step via gradient cosine similarity, enabling the scheduler to react immediately to changes in task relationships as training progresses.

\paragraph{RL Post-Training for Code LLMs.}
RL with verifiable execution rewards has become the dominant paradigm for post-training code LLMs~\cite{grpo, intro-RL-code-1}.
DeepSeek-R1~\cite{intro-RL-code-1} demonstrates that reinforcement learning with verifiable rewards can induce emergent reasoning behaviors such as self-reflection and verification in LLMs without requiring supervised rationale data.
Kimi k1.5~\cite{intro-RL-code-3} scales RL training with long-context rollouts and curriculum learning over problem difficulty.
RLEF~\cite{intro-RL-code-4} specifically targets code synthesis by training LLMs on multi-turn execution feedback, dramatically reducing the sample count needed relative to best-of-$N$ strategies.
CodeRL+~\cite{intro-RL-code-2} augments RL for code generation with execution semantics alignment, enabling models to reason about variable-level execution trajectories to achieve stronger generalization.
Lee et al.~\cite{intro-RL-code-5} apply adversarial RL to unit test generation, using an adversarial process to improve branch and statement coverage.
All of these approaches train task-specific specialists, and as shown in Figure~\ref{fig:performance}, each model degrades sharply outside its own training task.
{\tool} addresses this limitation by co-training a single model across all four coding tasks under a utility-guided multi-task RL framework.



\clearpage
\newpage

\section*{NeurIPS Paper Checklist}

\begin{enumerate}

\item {\bf Claims}
    \item[] Question: Do the main claims made in the abstract and introduction accurately reflect the paper's contributions and scope?
    \item[] Answer: \answerYes{}
    \item[] Justification: We have clearly stated the claims made in the abstract and introduction, accurately reflecting the paper’s contributions and scope.

\item {\bf Limitations}
    \item[] Question: Does the paper discuss the limitations of the work performed by the authors?
    \item[] Answer: \answerYes{}
    \item[] Justification: We have discussed the limitations of our work in the Appendix~\ref{sec:limit_future}, which include task and language coverage, reward quality for generative tasks, and the task-specific KL base coefficients.

\item {\bf Theory assumptions and proofs}
    \item[] Question: For each theoretical result, does the paper provide the full set of assumptions and a complete (and correct) proof?
    \item[] Answer: \answerNA{}
    \item[] Justification: Our paper does not include theoretical results, and therefore, this question is not applicable to our work.

    \item {\bf Experimental result reproducibility}
    \item[] Question: Does the paper fully disclose all the information needed to reproduce the main experimental results of the paper to the extent that it affects the main claims and/or conclusions of the paper (regardless of whether the code and data are provided or not)?
    \item[] Answer: \answerYes{}
    \item[] Justification: We provide a detailed description of the model and experimental settings in our paper, ensuring that readers have the necessary information to reproduce the main experimental results. Additionally, we plan to release the code to further enhance reproducibility and facilitate verification of our results.

\item {\bf Open access to data and code}
    \item[] Question: Does the paper provide open access to the data and code, with sufficient instructions to faithfully reproduce the main experimental results, as described in supplemental material?
    \item[] Answer:\answerNo{}
    \item[] Justification: We do not include the data and code with this version of the paper; we plan to release the code along with sufficient instructions to reproduce the main experimental results.

\item {\bf Experimental setting/details}
    \item[] Question: Does the paper specify all the training and test details (e.g., data splits, hyperparameters, how they were chosen, type of optimizer) necessary to understand the results?
    \item[] Answer: \answerYes{}
    \item[] Justification: We have provided all the necessary details regarding the training and testing process. Complete hyperparameter tables (including $\tau$, $\lambda$, task-specific KL base values, learning rate, etc.), baseline configurations, dataset splits, and evaluation metric definitions are provided in the Appendix.

\item {\bf Experiment statistical significance}
    \item[] Question: Does the paper report error bars suitably and correctly defined or other appropriate information about the statistical significance of the experiments?
    \item[] Answer: \answerNo{}
    \item[] Justification:  We conducted our experiments and baseline experiments on the same training and testing datasets to ensure a fair comparison.

\item {\bf Experiments compute resources}
    \item[] Question: For each experiment, does the paper provide sufficient information on the computer resources (type of compute workers, memory, time of execution) needed to reproduce the experiments?
    \item[] Answer: \answerYes{}
    \item[] Justification: We provided sufficient information on the computer resources needed to reproduce the experiments in the implementation details section.
    
\item {\bf Code of ethics}
    \item[] Question: Does the research conducted in the paper conform, in every respect, with the NeurIPS Code of Ethics \url{https://neurips.cc/public/EthicsGuidelines}?
    \item[] Answer: \answerYes{}
    \item[] Justification: We have reviewed the NeurIPS Code of Ethics. The paper studies multi-task reinforcement learning for code LLMs using publicly available or collected open-source datasets; it involves no human subjects and raises no specific ethical concerns.

\item {\bf Broader impacts}
    \item[] Question: Does the paper discuss both potential positive societal impacts and negative societal impacts of the work performed?
    \item[] Answer: \answerYes{}
    \item[] Justification: We discussed the social impact of this work in Appendix~\ref{sec:broader_impacts}, covering reduced deployment costs from unifying multiple specialists, and the dual-use risk of stronger automated code generation being misused for malicious purposes.
    
\item {\bf Safeguards}
    \item[] Question: Does the paper describe safeguards that have been put in place for responsible release of data or models that have a high risk for misuse (e.g., pre-trained language models, image generators, or scraped datasets)?
    \item[] Answer: \answerNA{} 
    \item[] Justification:  No such risks.

\item {\bf Licenses for existing assets}
    \item[] Question: Are the creators or original owners of assets (e.g., code, data, models), used in the paper, properly credited and are the license and terms of use explicitly mentioned and properly respected?
    \item[] Answer: \answerYes{}
    \item[] Justification: All datasets, benchmarks, and base models used in this work are publicly available and are cited in Section 3.1 and Appendix B; we use them in accordance with their licenses (e.g., Qwen2.5-Coder-7B and Qwen3-8B are released under the Apache 2.0 license).

\item {\bf New assets}
    \item[] Question: Are new assets introduced in the paper well documented and is the documentation provided alongside the assets?
    \item[] Answer: \answerNA{}
    \item[] Justification: This paper does not release new assets. We plan to release well-documented code for ASTOR.

\item {\bf Crowdsourcing and research with human subjects}
    \item[] Question: For crowdsourcing experiments and research with human subjects, does the paper include the full text of instructions given to participants and screenshots, if applicable, as well as details about compensation (if any)? 
    \item[] Answer: \answerNA{}
    \item[] Justification: Our research does not involve crowdsourcing nor research with human subjects.

\item {\bf Institutional review board (IRB) approvals or equivalent for research with human subjects}
    \item[] Question: Does the paper describe potential risks incurred by study participants, whether such risks were disclosed to the subjects, and whether Institutional Review Board (IRB) approvals (or an equivalent approval/review based on the requirements of your country or institution) were obtained?
    \item[] Answer: \answerNA{}
    \item[] Justification: Our research does not involve crowdsourcing nor research with human subjects.

\item {\bf Declaration of LLM usage}
    \item[] Question: Does the paper describe the usage of LLMs if it is an important, original, or non-standard component of the core methods in this research? Note that if the LLM is used only for writing, editing, or formatting purposes and does \emph{not} impact the core methodology, scientific rigor, or originality of the research, declaration is not required.
    \item[] Answer: \answerNA{}
    \item[] Justification: The core method development in our research does not involve LLMs as any important, original, or non-standard components.

\end{enumerate}

\end{document}